%% file: main.tex
\pdfoutput=1

\documentclass[journal]{IEEEtran}
\usepackage{ifpdf}

\usepackage{cite}

\ifCLASSINFOpdf
  \usepackage[pdftex]{graphicx}
\else
  \usepackage[dvips]{graphicx}
\fi
\usepackage{amsmath, amssymb}
\usepackage{algorithmic, algorithm}

\usepackage{array}

\ifCLASSOPTIONcompsoc
 \usepackage[caption=false,font=normalsize,labelfont=sf,textfont=sf]{subfig}
\else
 \usepackage[caption=false,font=footnotesize]{subfig}
\fi

\usepackage{stfloats}
\ifCLASSOPTIONcaptionsoff
 \usepackage[nomarkers]{endfloat}
\let\MYoriglatexcaption\caption
\renewcommand{\caption}[2][\relax]{\MYoriglatexcaption[#2]{#2}}
\fi
\usepackage{url}

\usepackage[colorlinks,linkcolor=red,anchorcolor=green,citecolor=blue]{hyperref}

\usepackage{multirow}
\usepackage{booktabs}
\usepackage{arydshln}
\usepackage[table,xcdraw,dvipsnames]{xcolor}
\usepackage{longtable} 
\usepackage{subfig}
\usepackage{subfloat}
\usepackage{booktabs}
\usepackage{threeparttable}
\usepackage{lscape}
\usepackage{threeparttable}
\usepackage{verbatim}
\usepackage{marvosym}
\usepackage{ulem}
\usepackage{fancyhdr}
\usepackage{soul}
\usepackage{siunitx}
\usepackage{pifont}

\makeatletter

\newcommand{\Rmnum}[1]{\expandafter\@slowromancap\romannumeral #1@}
\makeatother

\definecolor{c1}{HTML}{92a65f}
\definecolor{c2}{HTML}{F1BAA1}

\newcommand{\rOneCN}[1]{\textcolor{black}{#1}}
\newcommand{\rTwoCN}[1]{\textcolor{black}{#1}}
\newcommand{\rThreeCN}[1]{\textcolor{black}{#1}}
\newcommand{\rFourCN}[1]{\textcolor{black}{#1}}
\newcommand{\jOneFC}[1]{\textcolor{black}{#1}}
\newcommand{\WX}[1]{\textcolor{black}{#1}}

\newcommand{\reFirCN}[1]{\textcolor{black}{#1}}

\begin{document}

%
\title{SPEED: A \uline{S}calable RISC-V Vector \uline{P}rocessor \uline{E}nabling \uline{E}fficient Multi-Precision \uline{D}NN Inference}

%


\author{
    Chuanning~Wang, Chao~Fang, Xiao~Wu, Zhongfeng~Wang$^{(\textrm{\Letter})}$, Jun~Lin$^{(\textrm{\Letter})}$ \\
	\IEEEauthorblockA{
		School of Electronic Science and Engineering, 
		Nanjing University, China\\
		Email:
        \{zfwang, jlin\}@nju.edu.cn
    }
}

\author{Chuanning~Wang$^\dagger$,       Chao~Fang$^\dagger$,~\IEEEmembership{Graduate~Student~Member,~IEEE,}
        Xiao~Wu, \\
        Zhongfeng~Wang,~\IEEEmembership{Fellow,~IEEE}, 
        and~Jun~Lin,~\IEEEmembership{Senior~Member,~IEEE}
\thanks{$^\dagger$Equal contribution. This work was supported in part by the National Natural Science Foundation of China under Grant 62174084 and 62341408, and in part by the National Key R\&D Program of China under Grant 2022YFB4400604. \textit{(Corresponding author: Jun Lin and Zhongfeng Wang.)}}
\thanks{C. Wang, C. Fang, and X. Wu are with the School of Electronic Science and Engineering, Nanjing University (e-mail: chuan\_nwang@smail.nju.edu.cn; fantasysee@smail.nju.edu.cn; wxiao@smail.nju.edu.cn).}
\thanks{J. Lin is with the School of Electronic Science and Engineering, Nanjing University, and the Interdisciplinary Research Center for Future Intelligent
Chips (Chip-X), Nanjing University (email: jlin@nju.edu.cn).}
\thanks{Z. Wang is with the School of Electronic Science and Engineering, Nanjing University, and the School of Integrated Circuits, Sun Yat-sen University (email: zfwang@nju.edu.cn).}
}

%
%

\markboth{Journal of \LaTeX\ Class Files,~Vol.~14, No.~8, August~2015}%
{Shell \MakeLowercase{\textit{et al.}}: Bare Demo of IEEEtran.cls for IEEE Journals}
%



\maketitle

\input{0-abstr-r1.tex}

%
\IEEEpeerreviewmaketitle

\input{1-intro-r1.tex}
\input{3-design-r1.tex}

\input{4-res-r1.tex}
\input{5-concls.tex}
\ifCLASSOPTIONcaptionsoff
  \newpage
\fi



\normalem
\bibliographystyle{IEEEtran}
\bibliography{ref-full}
%




\end{document}

%% file: 0-abstr-r1.tex
\begin{abstract}
\WX{Deploying deep neural networks (DNNs) on those resource-constrained edge platforms is hindered by their substantial computation and storage demands.
Quantized multi-precision DNNs, denoted as MP-DNNs, offer a promising solution for these limitations but pose challenges for existing RISC-V processors due to complex instructions, suboptimal parallel processing, and inefficient dataflow mapping. 
To tackle the challenges mentioned above, SPEED, a scalable RISC-V vector (RVV) processor, is proposed to enable efficient MP-DNN inference, incorporating innovations in customized instructions, hardware architecture, and dataflow mapping. 
Firstly, some dedicated customized RISC-V instructions are introduced based on RVV extensions to reduce the instruction complexity, allowing SPEED to support processing precision ranging from 4-bit to 16-bit with minimized hardware overhead. 
Secondly, a parameterized multi-precision tensor unit is developed and integrated within the scalable module to enhance parallel processing capability \jOneFC{by providing reconfigurable parallelism that matches the computation patterns of diverse MP-DNNs.}
Finally, a flexible mixed dataflow method is adopted to improve computational and energy efficiency according to the \jOneFC{computing patterns of different DNN operators.}
The synthesis of SPEED is conducted on TSMC 28nm technology. Experimental results show that SPEED achieves a peak \reFirCN{throughput of 737.9 GOPS and an energy efficiency of 1383.4 GOPS/W }for 4-bit operators. 
Furthermore, SPEED exhibits superior area efficiency compared to prior RVV processors, with enhancements of  \reFirCN{5.9$\sim$26.9$\times$ and 8.2$\sim$18.5$\times$ for 8-bit operator and best integer performance}, respectively, which highlights SPEED’s significant potential for efficient MP-DNN inference.}
\end{abstract}


%% file: 1-intro-r1.tex
\section{Introduction}~\label{sec:intro}
\pdfoutput=1

\IEEEPARstart{D}{eep} \jOneFC{neural networks (DNNs) have demonstrated exceptional performance in complex tasks \cite{Dai2021DDETR, Nguyen2022Boxer, Cai2021UCSE, Lu2022LACS, Vaswani2017Attention, Brown2020LMFL, Tian2023Bebert, fang2022efficient} such as object detection, speech recognition, and natural language processing.
However, deploying these full-precision DNN models on edge devices faces significant challenges because of their high computational and memory requirements.
Multi-precision DNNs (MP-DNNs) address this issue by employing bit-widths smaller than the standard 32 bits to represent network parameters and activations~\cite{Wang2019CVPR, Neda2022ASPDAC, huang2024precision, zhou2018explicit}. 
By leveraging lower precision representations, such as 16, 8, or even 4-bit, these DNNs significantly reduce model size and computational demands, while maintaining near-equivalent accuracy to the full-precision counterparts~\cite{Rajagopal2018AccurateAE, Jain2018CompensatedDNNEE, Yin2019AnER}.
Hence, deploying MP-DNNs is crucial for optimizing performance and efficiency on resource-constrained edge platforms \cite{Wu2019FBNet, Li2022APE}.
}

\jOneFC{
In the pursuit of efficient MP-DNN deployment on edge devices, numerous specialized hardware accelerators have been proposed~\cite{Lee2018UNPU, Kang2021GANPU, Chen2016DianNao, Liu2016Cambricon, wu2021flexible}. 
However, these dedicated solutions face limitations in edge scenarios, mainly due to their limited flexibility and the complexity of configuration demands. 
While domain-specific hardware accelerators enhance execution efficiency for certain DNN models through fixed dataflows, they struggle to adapt to the rapidly evolving neural network architectures, diverse operators, and various inference precisions~\cite{Lee2018UNPU, Kang2021GANPU}.}
\jOneFC{In this case, RISC-V processor architecture~\cite{XpulpNN2021TETC, gautschi2017near, flamand2018gap, xu2022towards}, known for its ease of programming, presents a compelling approach to accommodate the flexible DNN workloads at the edge.}

\reFirCN{For instance, some works~\cite{XpulpNN2021TETC, Rossi2021VegaAT, Garofalo2023DARKSIDEAH, Dustin2023TCASI} extend the RISC-V instruction set with single-instruction-multiple-data (SIMD) instructions to accommodate MP-DNN tasks for achieving high energy efficiency. 
However, SIMD instructions lack the flexibility and scalability to accommodate various precision and tensor sizes, and thereby, some works require more instructions for the same task~\cite{XpulpNN2021TETC} and additional pipelines for fetching instructions~\cite{Rossi2021VegaAT, Garofalo2023DARKSIDEAH, Dustin2023TCASI}.}
In contrast, the RISC-V vector (RVV) extension instruction set architecture (ISA)~\cite{RVVExtension} is proposed as an effective solution, which features simple control mechanisms and flexible scalability. 
Compared to the RISC-V base ISA~\cite{Waterma2016EECS}, \reFirCN{the RVV ISA enables the multi-precision operation of vectors with a configurable length by a single instruction to enhance flexibility of parallel processing, making it particularly well-suited for MP-DNN inference in edge computing scenarios.}
\jOneFC{Despite the advancements of the RVV extensions, challenges persist in deploying MP-DNNs on RVV-based processors~\cite{Cavalcante2022NewAra, Perotti2023Yun, Askarihemmat2023QuarkAI, Theo2023sparq}.
The official RVV ISA supports operations from 8-bit to 64-bit, yet lacks native handling for low-precision, such as 4-bit, which is crucial for MP-DNNs.} 
\reFirCN{While some RVV processors~\cite{Askarihemmat2023QuarkAI, Theo2023sparq} mitigate this by introducing customized instructions for low-precision computing, this approach lacks the consideration of the practical applications for MP-DNN inference due to the notable loss of accuracy.
Moreover, the potential of deploying models at 4-bit precision, which could enhance computational throughput without compromising inference accuracy, still needs to be explored.}

Furthermore, \reFirCN{some RISC-V processors with cluster architecture combine different single-precision modules for multi-precision computation without reusing, increasing hardware resource consumption~\cite{XpulpNN2021TETC, Garofalo2023DARKSIDEAH, Dustin2023TCASI}.}
The hardware architecture of RVV processors often features scalable modules that only enable parallelism along a single dimension~\cite{Cavalcante2020AraA1TVLSI, Askarihemmat2023QuarkAI, Theo2023sparq, Cavalcante2022NewAra, Perotti2023Yun}.
\jOneFC{This limitation hampers the full exploitation of data reuse in multi-dimensional tensor computations,} \WX{particularly in convolution operations, leading to a degradation of hardware performance.}

\jOneFC{Additionally, current limitations in dataflow mapping hinder the efficiency of these processors.}
\jOneFC{By using a uniform dataflow mapping method across all DNN operators, including standard convolution (CONV) and depth-wise convolution (DWCV), these processors fail to adapt to the varying compute-storage demands of different operations, leading to computational inefficiency and increased off-chip communication~\cite{Cavalcante2020AraA1TVLSI, Cavalcante2022NewAra}.}
\jOneFC{In summary, deploying MP-DNNs on \reFirCN{RISC-V} processors still struggles with 1) limited support of RISC-V ISA for \reFirCN{practical MP-DNN applications}, 2) \reFirCN{multi-precision computation and single-dimensional parallelism bottleneck of hardware} architecture, and 3) the suboptimal dataflow mapping for various DNN operators.}
\WX{To address the issues above, we propose SPEED, a scalable RVV processor that enables efficient MP-DNN inference across 4, 8, and 16 bits. 
By incorporating customized vector instructions based on the official RVV ISA,  an efficient hardware architecture, and a flexible dataflow mapping strategy, SPEED aims to improve the area efficiency and inference throughput of MP-DNN deployments within the constraints of edge computing scenarios.}
The contributions of this work can be summarized in the following aspects:
\begin{enumerate}
    \item[\textbf{1)}] \textbf{Customized instructions}: 
    \WX{
    Multiple customized vector instructions are proposed by utilizing the reserved user-defined encoding space within the official RISC-V ISA, \reFirCN{optimizing configuration-setting, data access, and DNN operator deployment.}
    These instructions \reFirCN{enable SPEED to support multi-precision operations flexibly and} are optimized for specific operations in MP-DNN inference, further reducing the instruction count required compared to the official ISA.}
    \item[\textbf{2)}] \textbf{Hardware architecture}: 
    \WX{A scalable RVV processor, namely SPEED, is specifically designed for efficient MP-DNN inference. 
    It integrates a \reFirCN{configurable and programmable} multi-precision tensor unit \reFirCN{that enables parallelism across multi-dimension} inside scalable modules to expand the range of parallelism options for multiple operators, significantly enhancing the parallel processing capability of SPEED.
    Compared to prior arts~\cite{Perotti2023Yun, Rossi2021VegaAT, XpulpNN2021TETC, Garofalo2023DARKSIDEAH, Dustin2023TCASI}, our design achieves an improvement} 
    \reFirCN{in area efficiency by 5.9$\sim$26.9$\times$ and 8.2$\sim$18.5$\times$ for 8-bit and best integer performance conditions, respectively.}
    \item[\textbf{3)}] \textbf{Dataflow mapping}: 
    \jOneFC{A mixed dataflow mapping method, encompassing a variety of strategies, is designed to align with the specific computing and storage requirements of different DNN operators, thereby enhancing hardware flexibility and computational efficiency.}
    \WX{By employing these optimized dataflow strategies, our design can maximize data reuse, significantly reducing execution latency and the size of off-chip transferred data.}
    \reFirCN{
    Benefiting from the proposed mixed dataflow mapping method, SPEED improves average throughput by 4.8$\times$ and 11.8$\times$ over Ara~\cite{Cavalcante2022NewAra} when evaluated on DNN benchmarks at 16-bit and 8-bit precisions, respectively.
    }
\end{enumerate}

\jOneFC{
The remainder of this paper is organized as follows.
Sec.~\ref{sec: Archi_design} elaborates on the hardware innovations of SPEED from customized instructions and hardware architecture. 
Sec.~\ref{sec: dataflow_design} presents the scheduling innovations of SPEED from the flexible dataflow mapping.
Sec.~\ref{sec:res} demonstrates the superior performance of SPEED with extensive experimental results.
Sec.~\ref{sec:concls} concludes the paper.
}

%% file: 3-design-r1.tex
\pdfoutput=1




\section{The Proposed SPEED Architecture}\label{sec: Archi_design}
\jOneFC{SPEED is a scalable RVV processor enabling area-efficient and high-throughput MP-DNN inference, built upon the RVV v1.0 ISA. 
It is tightly coupled to a RISC-V scalar core for flexible programming and equipped with an external memory for efficient data retrieval. 
SPEED is designed to address the evolving challenges of MP-DNN inference by integrating specialized instructions with an efficient hardware architecture, and this section will detail how these two innovations enhance MP-DNN inference efficiency.
}
\begin{figure}[!htbp]
  \centering
    \includegraphics[width=0.88\columnwidth]{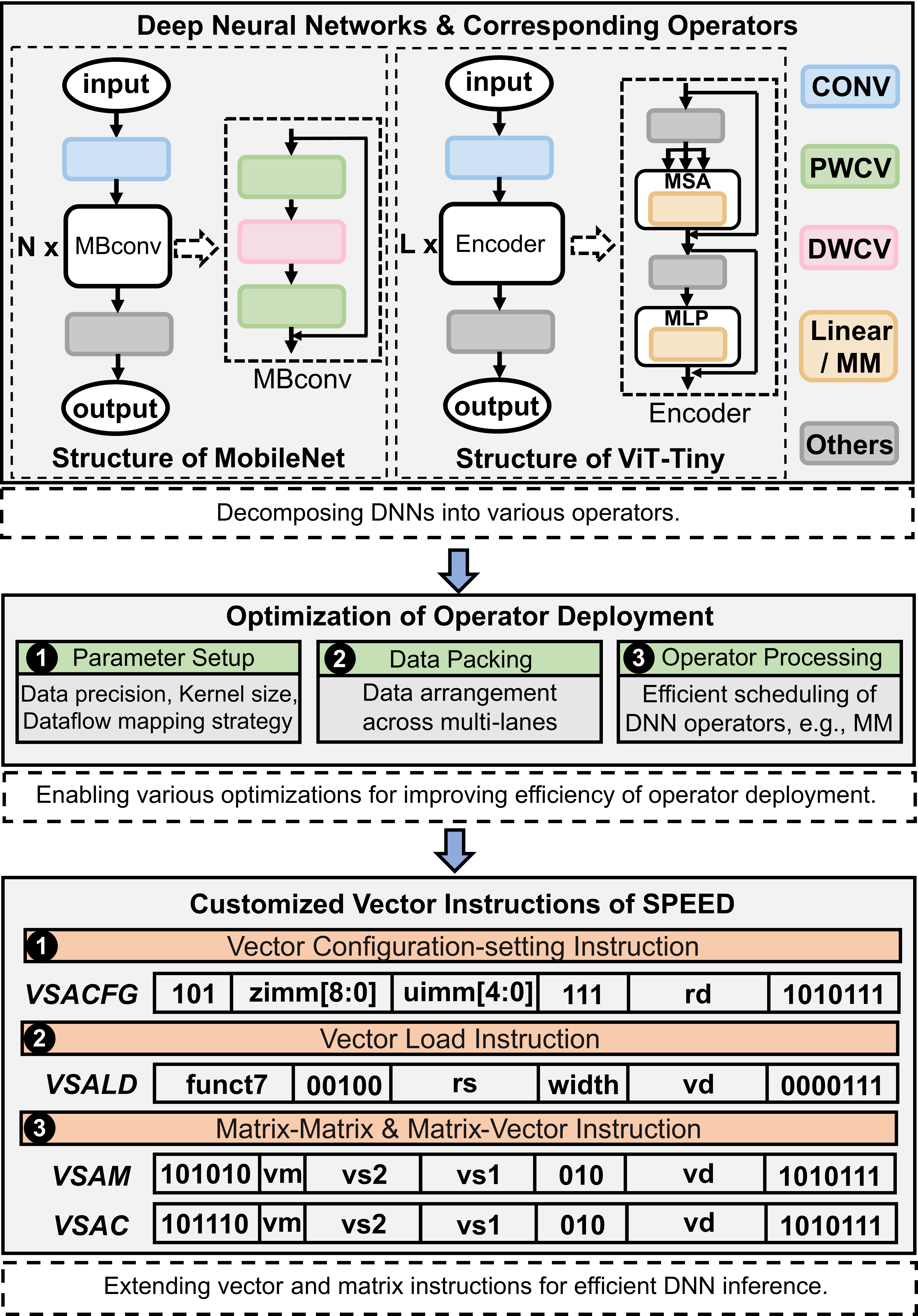}
      \caption{\jOneFC{Customized vector instructions of SPEED drive enhanced computational efficiency in operators for MP-DNN inference.}}
        \label{fig:customized_instr}
\end{figure} 

\subsection{Motivations for Customized Vector Instructions}\label{subsec: motivations}
\jOneFC{While the RVV ISA extension~\cite{RVVExtension} is powerful for general-purpose computing, it encounters some limitations when being leveraged to perform MP-DNN inference~\cite{Zhao2023HIPU}.}
As shown in Fig.~\ref{fig:customized_instr}, MP-DNNs can be decomposed into various operators.
\jOneFC{Note that convolution operators in the popular convolutional neural networks (CNNs) account for more than 95\% of the total inference time~\cite{PCONV2020AAAI}, which include standard convolution (CONV), point-wise convolution (PWCV), and depth-wise convolution (DWCV)~\cite{Liu2015VGG16, He2016ResNet, Sandler2018MobileNet}, constituting the majority of the computational workload in CNNs.}
\jOneFC{Similarly, matrix multiplication (MM) operations in Transformer models~\cite{dosovitskiy2021an, Yang2019XLNet, Rao2021DynamicViT, fang2022algorithm} are computationally expensive, accounting for a significant portion of the model's parameters and computations~\cite{GPT32022NeurIPS}.}
\jOneFC{However, processors relying solely on the official RVV instruction set, such as Ara~\cite{Cavalcante2022NewAra}, struggle to efficiently handle the unique computational characteristics of these operators~\cite{Zhao2023HIPU}.}
\jOneFC{The arising challenges include limited supported precisions, a high instruction count, and the inefficiency in leveraging the reuse opportunities of scalable modules, which not only lead to decoding delays that impair overall computational efficiency but also increase hardware resource consumption due to the complexity of the decoding units.}

To address the limitations above, SPEED introduces several optimization techniques when designing customized RVV instructions as illustrated in Fig.~\ref{fig:customized_instr}. \jOneFC{These optimizations aim to simplify operator parameter configuration, enhance data reuse across scalable modules, and directly compute specific operator functions, thereby improving the overall efficiency of MP-DNN inference.}

\subsection{Customized Efficient Vector Instructions of SPEED}\label{subsec: instr_design}

\rTwoCN{Based on the analysis of operator deployment for efficient MP-DNN inference, as shown in Fig.~\ref{fig:customized_instr}, several customized instructions are proposed using the reserved user-defined encoding space in RISC-V.}
\rThreeCN{Specifically, the customized instructions mainly contain the configuration-setting (\texttt{VSACFG}), memory access (\texttt{VSALD}), and arithmetic (\texttt{VSAM}, \texttt{VSAC}).}

\texttt{VSACFG} serves as an efficient vector configuration-setting instruction that effectively provides the necessary information of specific operators, such as data precision (4-, 8-, and 16-bit), 
\reFirCN{size of convolution kernel (1$\sim$15),} 
and dataflow mapping strategy.
The above information are decoded according to the value of $zimm$[8:0] and $uimm$[4:0], respectively, and then specified in a control register ($rd$) for the subsequent computations, as shown in Fig.~\ref{fig:customized_instr}.
\reFirCN{
For convolution computations with a kernel size larger than 15,
inspired by the Kseg method in work~\cite{Wu2024Amoeba}, the larger kernels are decomposed into several smaller sub-kernels according to our computational parallelism, thus saving the storage resources for processing DNN operations.
}

\texttt{VSALD} is a customized vector memory access instruction designed to efficiently load data from the external memory. 
\jOneFC{Unlike the official RVV load instruction \texttt{VLE} that performs sequential allocation, \texttt{VSALD} facilitates data reuse by enabling simultaneous loading and multi-broadcasting of data from external memory to individual scalable modules.}
The vector load instruction \texttt{VSALD} is as follows:
$$ \texttt{VSALD(width).v\quad vd, (rs)}\text{,} $$
where the vector load moves data from external memory with address (\texttt{rs}) to the vector register \texttt{vd} at the precision which is determined by \texttt{width}. 

\texttt{VSAM} and \texttt{VSAC} are customized arithmetic instructions that exploit data parallelism for matrix-matrix operations and matrix-vector operations, respectively.
\jOneFC{Unlike the arithmetic instructions in prior works~\cite{Cavalcante2022NewAra, Cavalcante2020AraA1TVLSI, Perotti2023Yun} that are limited by a single parallelism dimension, the proposed \texttt{VSAM} and \texttt{VSAC} can directly perform operations on entire matrices or vectors within multiple parallelism dimensions.}
\jOneFC{This eliminates the need to decompose operations into multiple smaller ones, significantly reducing the number of instructions, data movements, and required vector registers.}
\jOneFC{Hence, \texttt{VSAM} and \texttt{VSAC} offer improved computational efficiency and lower resource consumption.}
The \texttt{VSAM} and \texttt{VSAC}  vector arithmetic instructions are as follows:
$$ \begin{cases}\texttt{VSAM.vv\quad vd,\,vs1,\,vs2}\text{,} \\ \texttt{VSAC.vv\quad vd,\,vs1,\,vs2}\text{,}\end{cases}$$
\jOneFC{where \texttt{vs1} and \texttt{vs2} are source vector registers holding the input operands. The computation result is stored in the destination vector register, denoted as \texttt{vd}, enabling efficient post-processing with vector instructions on chip and eliminating the need for additional off-chip communication.}


\reFirCN{
By effectively organizing the above-customized instructions, various DNN workloads can be efficiently implemented on the processor.
At the software level, these customized instructions, combined with official RISC-V vector instructions, can be written using inline assembly code within C/C++ programs and compiled with the developed RISC-V GNU/LLVM toolchain~\cite{GNUToolchain, LLVMToolchain} for program debugging and assembly code generation.
On the hardware level, SPEED efficiently decodes vector instructions within the compiled program to access the necessary execution information for performing the corresponding operations.
}
\begin{figure}[!htbp]
  \centering
    \includegraphics[width=0.92\columnwidth]{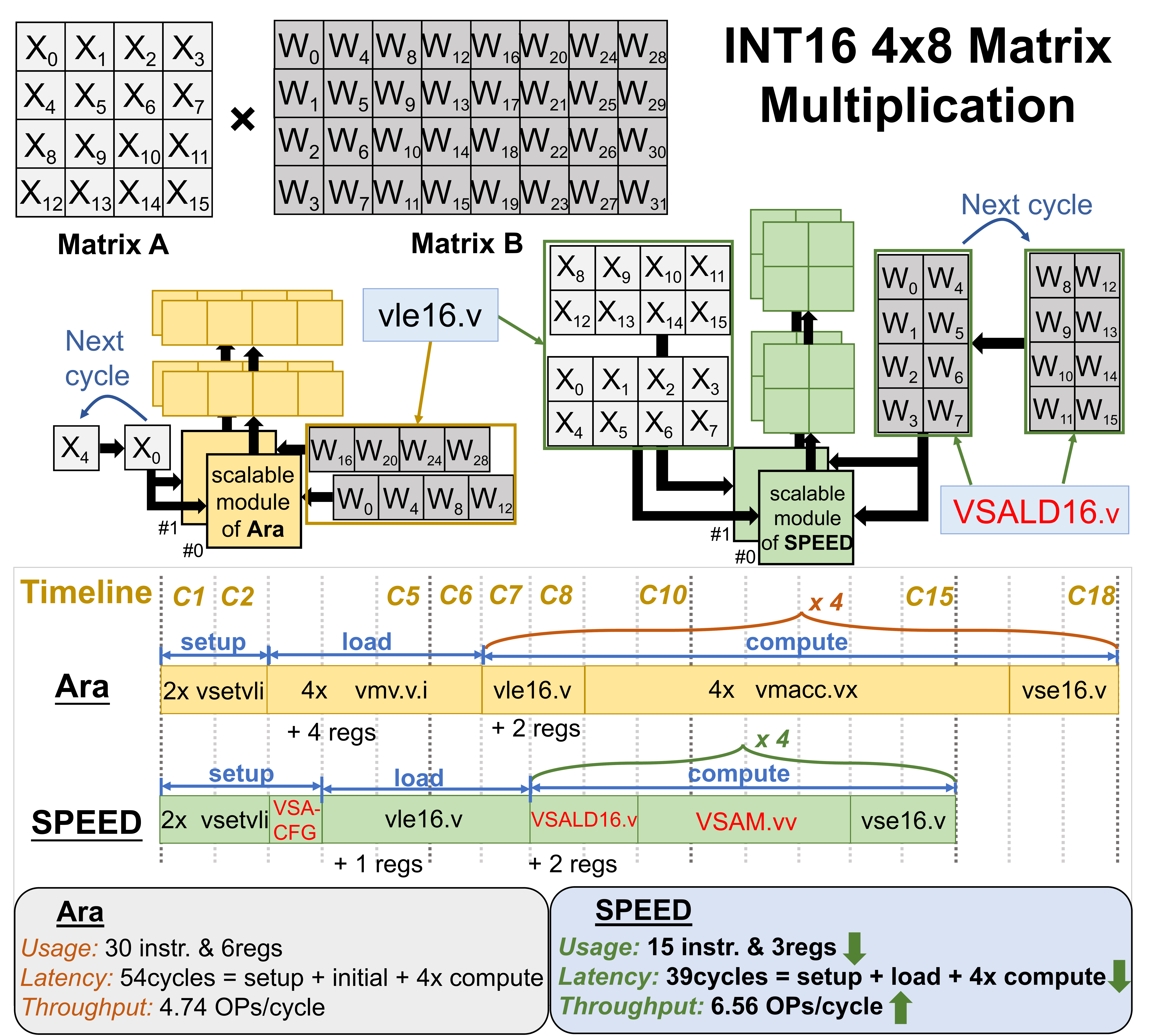}
      \caption{\reFirCN{Comparison of SPEED and Ara when performing an INT16 MM operator. SPEED leverages fewer instructions, requires smaller number of vector registers, and achieves fewer computing cycles over Ara.}}
        \label{fig:instr_compare}
\end{figure}

\jOneFC{To \reFirCN{explain the programmability of SPEED} and demonstrate the advantages of customized vector instructions, we compare the processing instruction sequences of SPEED and Ara for a 4$\times$8 MM operator at 16-bit precision with a focus on the processing cycles and register usage for both architectures.}
\reFirCN{As shown in Fig.~\ref{fig:instr_compare}, both SPEED and Ara involve setup, load and compute stages for parameter configuration and execution, respectively.}
\reFirCN{The setup stage configures parameters for the subsequent computation, such as 16-bit execution precision.}
\rOneCN{Specifically, in the setup stage, both processors utilize the \texttt{VSETVLI} instruction to set the application vector length, managing computational resources for vector operations. SPEED, however, introduces the \texttt{VSACFG} instruction, enabling direct configuration of parameters crucial for matrix and convolution computations.}
\reFirCN{During the load stage, SPEED and Ara use different memory access instructions to fetch required computational data, with SPEED utilizing fewer registers in this process.}
\jOneFC{Moving to the compute stage, SPEED leverages the \texttt{VSALD} instruction to achieve multi-broadcasting of data across scalable modules, facilitating efficient data reuse.}
\jOneFC{In addition, as shown in Fig.~\ref{fig:instr_compare}, Ara relies on 16 individual arithmetic instructions, namely \texttt{VMACC},
\reFirCN{for computations under 16-bit precision,} 
whereas SPEED achieves the same result with only four \texttt{VSAM} instructions.}
\rThreeCN{Finally, both Ara and SPEED employ four \texttt{VSE} instructions to store the computation results.}
\jOneFC{By employing customized vector instructions tailored to specific operators, SPEED achieves a significant efficiency improvement compared to Ara. 
In this case, SPEED utilizes 46\% fewer instructions and 50\% fewer registers, leading to a 1.4$\times$ throughput boost over Ara.}


\subsection{Micro-architecture of SPEED}

\begin{figure}[!htbp]
  \centering
    \includegraphics[width=1\columnwidth]{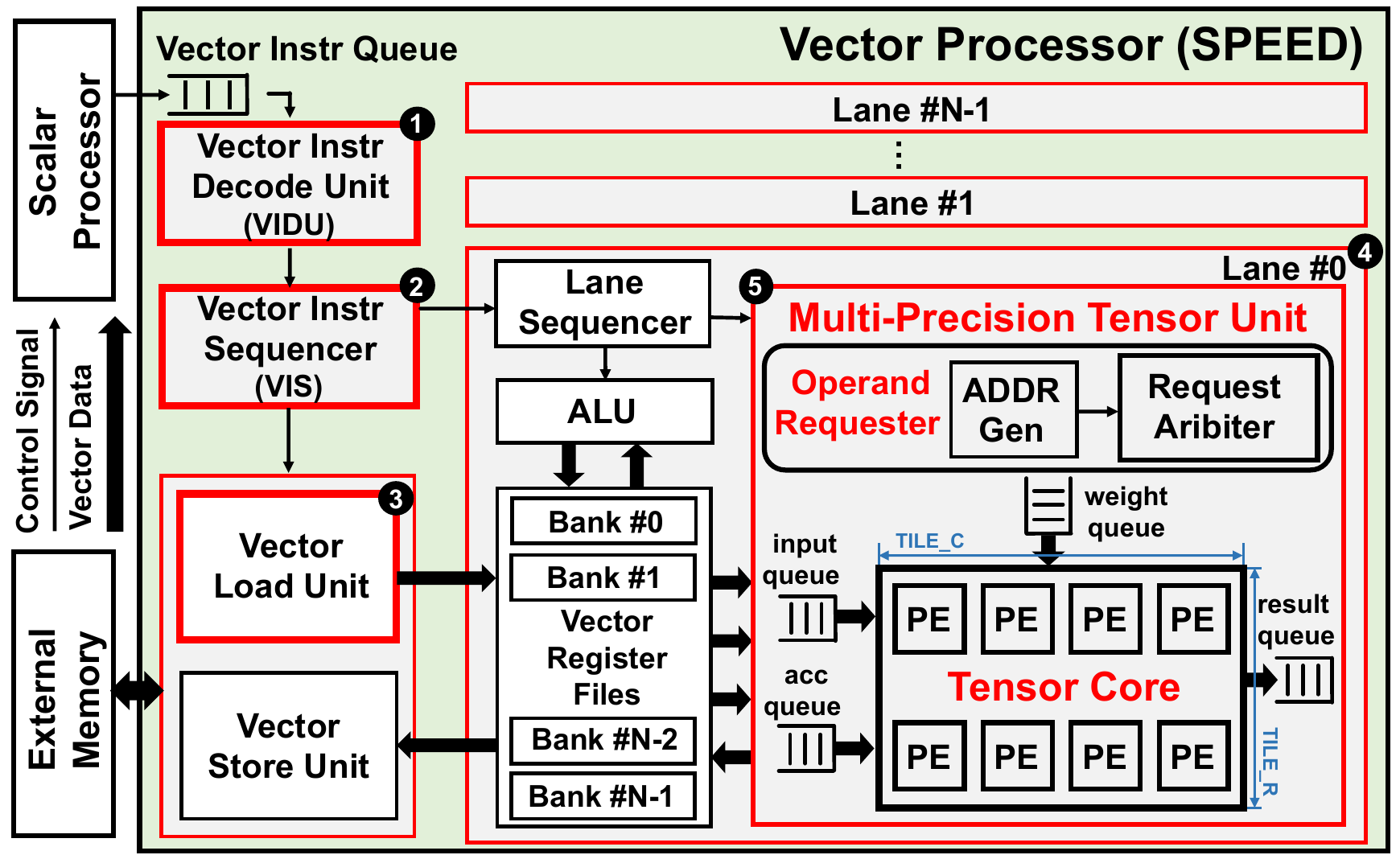}
      \caption{\reFirCN{Micro-architecture of SPEED, the proposed RVV processor.}}
        \label{fig:micro_arichi}
\end{figure}

\reFirCN{SPEED is a 4-stage pipeline vector processor that implements official RVV ISA, plus customized multi-precision instructions operating on vector elements.
}
\WX{The detailed SPEED micro-architecture is shown in Fig.~\ref{fig:micro_arichi}, containing a \ding{202}~vector instruction decode unit (VIDU), a \ding{203}~vector instruction sequencer (VIS), a \ding{204}~vector load unit (VLDU), a vector store unit, and multiple parallel computing \ding{205}~lanes, which can efficiently support customized vector instructions to enhance the overall MP-DNN inference performance.} 

\reFirCN{Specifically, \ding{202}~VIDU is developed to decode all vector instructions from scalar processor for efficient MP-DNN inference, which involves official RVV and customized vector instructions.
VIDU also propagates the execution results of vector instructions to the scalar processor.}
\reFirCN{The internal register \textit{rd} of VIDU is accountable for storing the execution precision information within the decoded information of \texttt{VSACFG} instruction.}
\reFirCN{\ding{203}~VIS handles these decoded instructions and sends them to the appropriate functional units (FUs).
VIS is also responsible for keeping track of the vector instructions running on SPEED and storing the information about which vector register is accessed by operating instructions, aiming to avoid the hazard of local memory access among different instructions.}
\jOneFC{Additionally, \ding{204}~VLDU enables efficient data movement, offering sequential transfer and multi-broadcast capabilities from external memory to scalable modules.}
\WX{The multi-mode VLDU design allows our SPEED processor to meet the requirements of diverse data reuse patterns for various operators.}
Moreover, the scalable modules of SPEED, namely \ding{205}~lane, serve as the main computational components.
\jOneFC{Each lane is equipped with a suite of components that \WX{perform} efficient data management and computation, including a lane sequencer, vector register files (VRFs), an arithmetic logic unit (ALU), and a multi-precision tensor unit (MPTU).
The lane sequencer is responsible for requesting data reads from VRFs and responding to VIS upon completing instruction executions.
The VRFs within each lane serve as the local memory, enabling rapid access to data involved in computations.
The ALU performs general arithmetic operations\WX{, while the proposed MPTU focuses on} executing computation-intensive multi-precision tensor operations of MP-DNNs with high efficiency.}

\subsection{Hierarchical Structure of MPTU}
\jOneFC{To achieve efficient multi-precision parallel data processing and maximize data reuse, \reFirCN{a programmable and configurable MPTU} is incorporated into SPEED's lane as the core computing element.}
\reFirCN{The MPTU is controlled by customized instructions \texttt{VSAM}, \texttt{VSAC} and comprises an operand requester, queues, and a tensor core.}

The operand requester consists of an address generator and a request arbiter, enabling efficient operand access by concurrently generating addresses and prioritizing operand requests. 
The queues are responsible for buffering the operands obtained through operand requests, including inputs, weights, accumulation data, and results.
\begin{figure}[!tp]
    \centering
    \includegraphics[width=0.92\columnwidth]{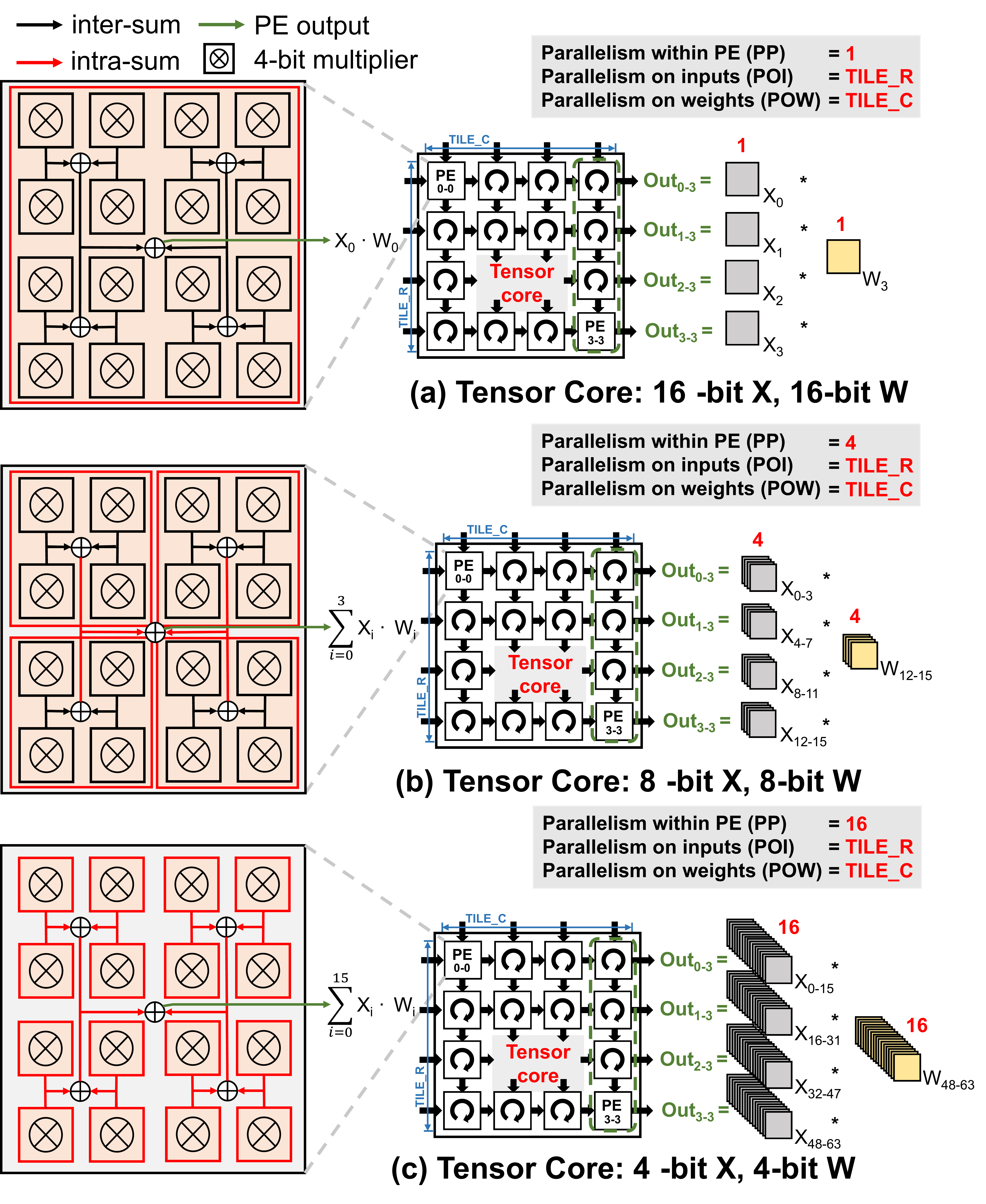}
    \caption{\reFirCN{Hierarchical architecture of multi-precision tensor core and exploited data parallelisms under 16-bit, 8-bit, and 4-bit data precision, respectively.}}
    \label{fig:mult_precision_SA}
\end{figure}

\jOneFC{The tensor core is a two-dimensional array of processing elements (PEs) configurable by \#TILE\_R and \#TILE\_C parameters, which is designed to adapt to the throughput requirements of diverse MP-DNN applications.}
\reFirCN{As illustrated in Fig.~\ref{fig:mult_precision_SA}, the tensor core maximizes the reuse of inputs and weights by efficiently organizing the internal dataflow, and each PE employs an output-stationary execution strategy to utilize partial sums, collectively enhancing computation efficiency effectively.}
\WX{The decision of whether to temporarily retain the 32-bit output within the PE or transfer it horizontally to the result queue depends on the tensor core's state, specifically, the completion of the current computation.}
\WX{Each PE stores the incoming data in an internal register and forwards the same data to its neighbor PE.
These store and forward behaviors contribute to significant savings in external memory accesses, effectively exploiting data reuse opportunities.}
As depicted in Fig.~\ref{fig:mult_precision_SA}, the proposed PE contains sixteen 4-bit multipliers, which can perform one set of 16-bit multiply-accumulate (MAC) \WX{operations}, four sets of 8-bit MAC operations, and sixteen sets of 4-bit MAC operations.


Fig.~\ref{fig:mult_precision_SA} shows that tensor core architecture employs three levels of parallelism, \WX{denoted by} parallelism within PE ($PP$), parallelism on inputs ($POI$), and parallelism on weights ($POW$). 
The value of $PP$ depends on the operand precision. For 16-bit, 8-bit, and 4-bit operands, $PP$ is 1, 4, and 16, respectively.
Meanwhile, the values of $POI$ and $POW$ are determined by the size of the tensor core, i.e., \#TILE\_R and \#TILE\_C.
\WX{The parallel computation within PE is implemented to minimize the extra data movement provided by overlapping data between adjacent computations, fully leveraging data reuse opportunities.
The parallel computing data for inputs and weights are arranged in an orthogonal pattern to align with the structure of the multi-dimensional array, significantly enhancing the computational efficiency of the MPTU.}
\WX{Moreover, utilizing these three levels of parallelism can offer a flexible dataflow mapping opportunity, enabling the tensor core with high computational efficiency for various MP-DNN  operators, which is elaborated in Sec.~\ref{sec: dataflow_design}.}

\subsection{Pipeline Stage Allocation and Runtime Reconfigurability}
\begin{figure}[!tb]
  \centering
    \includegraphics[width=0.94\columnwidth]{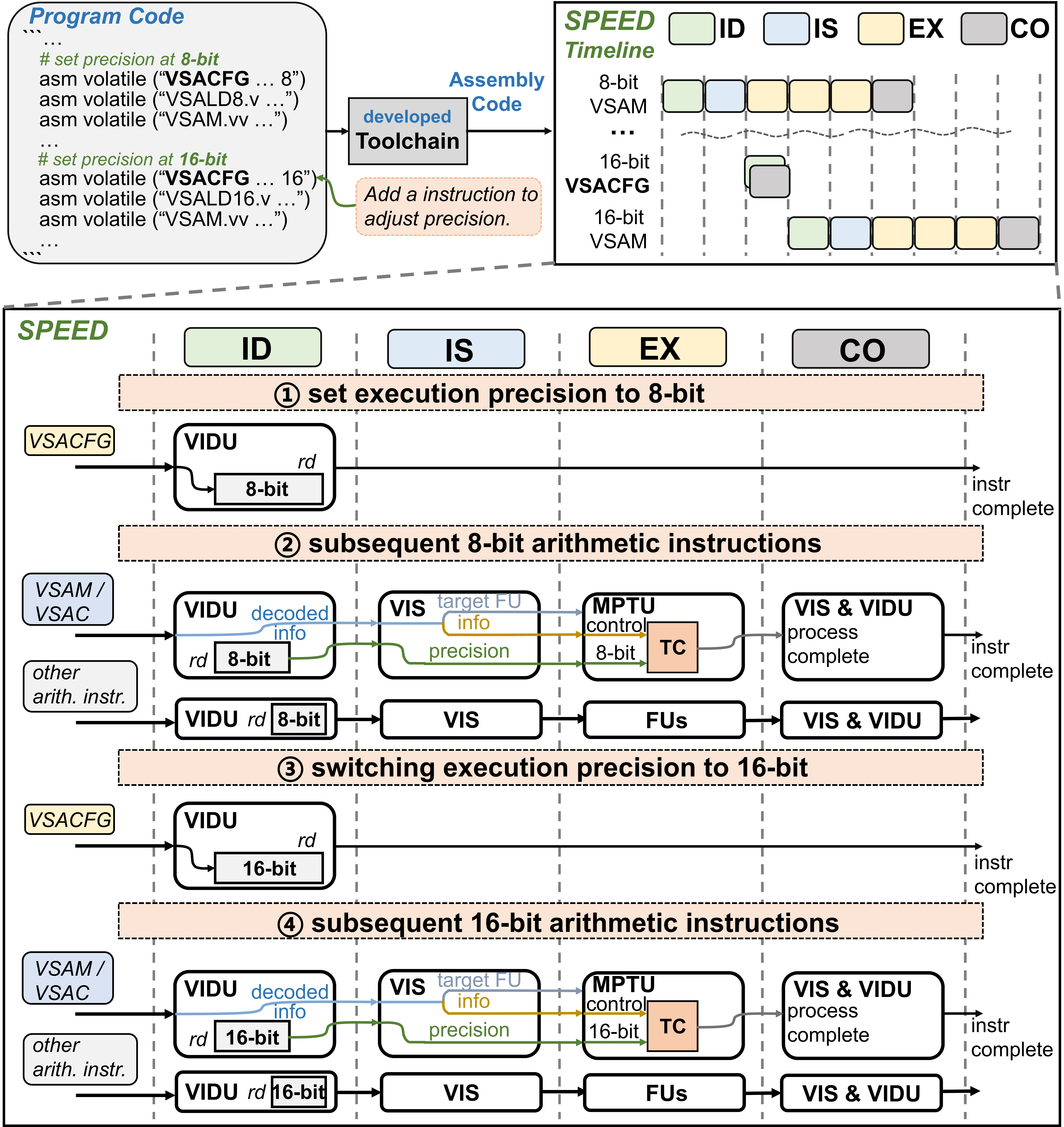}
      \caption{\reFirCN{The pipeline stages allocation and operation precision switching method in the proposed design.}}
        \label{fig:instr_pipeline}
\end{figure}

\reFirCN{
To gain a comprehensive understanding of how SPEED allocates pipeline stages and enables runtime precision reconfigurability, we present an example of executing a program involving both 8-bit and 16-bit precision operations on SPEED.
The pipeline in SPEED is organized into four stages: Instruction Decode (ID), Instruction {iSsue} (IS), {EXecution} (EX), and {COmmit} (CO). 
As depicted in Fig.~\ref{fig:instr_pipeline}, the program involving vector instructions is compiled by the developed toolchain~\cite{GNUToolchain, LLVMToolchain}. 
Then, SPEED operates these vector instructions through the 4-stage pipeline to perform the corresponding operations.}

\reFirCN{
During the ID stage, the VIDU decodes the vector instructions for obtaining the execution information, and the internal register $rd$ stores the execution precision information.
Specifically, the 8-bit precision information in the current \texttt{VSACFG} instruction is decoded and stored in the $rd$ of the VIDU, as shown in step \ding{172}.
In the IS stage, the VIS recognizes the target FUs of these instructions. 
The precision and decoded information are then sent to the recognized FUs.
For example, with customized vector instructions \texttt{VSAM} and \texttt{VSAC}, as shown in step~\ding{173} of Fig.~\ref{fig:instr_pipeline}, the 8-bit precision information is synchronized with the decoded information of the instruction in the IS stage and then issue to the MPTU.
For the EX stage, FUs such as MPTU and VLDU perform specific operations according to the received precision information and other information (e.g., storage locations of involved operands).
As depicted in Fig.~\ref{fig:instr_pipeline}, the operating precision of MPTU is configured by the specific execution precision information attached to the decoding of arithmetic instructions.
The cycle counts in the EX stage are determined by the workload's tensor size, unlike the ID and IS stages, which only require one cycle.
Once the executions are completed, the current operating instruction enters the CO stage.
At the CO stage, the VIS cancels the execution status of the current arithmetic instruction and clears the register occupancy indicator. 
At the same time, the VIDU propagates the instruction finish signal to the scalar core.
Notably, as illustrated in Fig.~\ref{fig:instr_pipeline}, the \texttt{VSACFG} instruction involves ID and CO stages and allows flexible precision conversion with a single cycle once the \textit{rd} register in the VIDU is updated.}

\reFirCN{
SPEED enables runtime configurability of execution precision by incorporating specific configuration-setting instructions, as shown in Fig.~\ref{fig:instr_pipeline}.
When switching execution precision at runtime, a new configuration-setting instruction should be implemented before related instructions to adjust the execution precision accordingly, only within one cycle.
For instance, by using a configuration-setting instruction to update the 8-bit precision information in the VIDU to 16-bit, subsequent instructions will synchronize with the 16-bit precision and be sent to the appropriate FUs for processing, as demonstrated in steps \ding{174} and \ding{175}.}
\reFirCN{
By implementing precise pipeline stage allocation, SPEED attains a high operating frequency, while the low-latency runtime precision switching method enhances the computational efficiency of multi-precision operations, leading to effective and high-performance MP-DNN inference.
}

\section{The Proposed Flexible Mixed Dataflow}\label{sec: dataflow_design}
\WX{
\jOneFC{This section elaborates on the proposed flexible mixed dataflow mapping method, which aims to optimize the data arrangement for MP-DNN computing.}
\jOneFC{It consists of a matrix multiplication dataflow strategy for the MM operator, along with Feature-Map-First-Channel-Second~(FFCS), Channel-First~(CF), and Feature-Map-First~(FF) strategies for the CONV, PWCV, and DWCV operators, respectively.}
These strategies fully exploit data reuse opportunities brought by these operators and computation parallelism in SPEED, further improving MP-DNN deployment performance.
}
\begin{figure*}[!htbp]
  \centering
    \includegraphics[width=1.6\columnwidth]{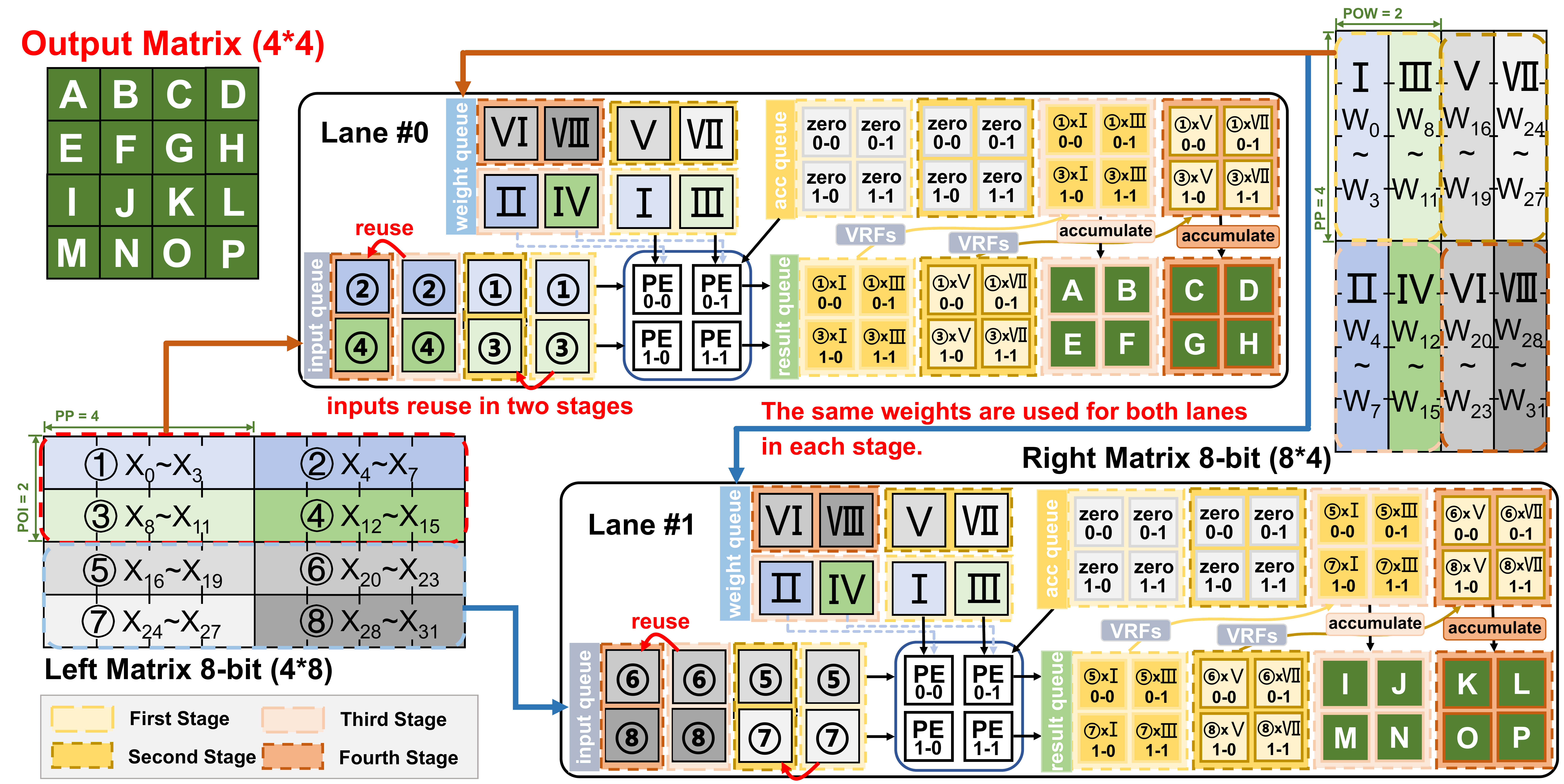}
      \caption{\reFirCN{An example of computing dataflow strategy for 8-bit MM operators.}}
        \label{fig:matu_dataflow}
        \vspace{-10pt}
\end{figure*}

\subsection{Matrix Multiplication Dataflow Strategy}\label{sec: matrix_dataflow}
\WX{MM is a fundamental operation that forms the core computation within Transformer. 
Additionally, convolution operations can be converted into MM operators for efficient processing. 
Thus, optimizing \jOneFC{the scheduling of MM operators} becomes crucial for improving the performance of MP-DNN deployments.}
\jOneFC{Limited by its three computation loops, the MM operator suffers from restricted data reuse and parallelism.}
\WX{To solve this issue, we propose an efficient MM dataflow strategy to perform the parallel MM execution.}
\jOneFC{Specifically, it leverages multi-broadcasting of weights across scalable modules and input reuse across processing stages, optimizing the arrangement of parallel computing data and achieving significant gains in computational efficiency.}

\WX{To elaborate on the proposed MM dataflow strategy, we take the four processing stages of 8-bit MM operators as an example, depicted in Fig.~\ref{fig:matu_dataflow}.}
\WX{The processor is configured with two lanes, where each lane is equipped with a tensor core of size 2$\times$2 to satisfy the requirements of parallel computations ($POI$ = $POW$ = 2).
Within each lane, the data in $POI$ adjacent rows of the left matrix are multiplied simultaneously with those in $POW$ adjacent columns. Additionally, $PP$ data from adjacent computations are processed in the same PE to fully exploit the parallelism provided by MPTU.
}

\WX{
As shown in Fig.~\ref{fig:matu_dataflow}, the input queue and weight queue of each lane send computing data to PEs, where the results are organized in the result queue and subsequently stored in the VRFs.
Considering the $PP$ is set as 4 in 8-bit computations, four neighbouring data of the computation matrix are combined as a single operand, such as inputs~\ding{192}. 
In the first stage, inputs \ding{192}, \ding{194} and \ding{196}, \ding{198} are allocated to two lanes, respectively, with each lane reusing the same weights \Rmnum{1} and \Rmnum{3}. 
The weight queue requests weights \Rmnum{5} and \Rmnum{7} for the next stage, while the same data in the input queue are reused in PEs to perform computations along the row dimension of the right matrix.
In the third stage, the input queue fetches new data (inputs \ding{193}, \ding{195} and \ding{197}, \ding{199}) along the rows of the left matrix for computation in each lane by reusing weights \Rmnum{2} and \Rmnum{4}. 
Besides, the results generated in the first stage, temporarily buffered in the accumulation queue to reduce the off-chip data transfer, are requested from the VRFs during the computation process. 
They are added to the results of the current stage to obtain output results. 
Similarly, in the fourth stage, the same data in the input queue from the previous stage are multiplied by their corresponding weights \Rmnum{6} and \Rmnum{8}. 
The results are accumulated with the partial sum from the second stage to complete the computation of one row in the output matrix.}

\subsection{Convolution Dataflow Strategy}\label{subsec: conv_dataflow}
\jOneFC{Convolution operators are critical for the performance of CNNs.
\WX{In pursuit of higher performance, convolutional variants like DWCV and PWCV have been developed to satisfy the demands of various tasks on top of CONV.}
However, their unique computational and storage needs mean that a one-size-fits-all dataflow approach would suffer from under-utilized computation.}
\WX{To solve this issue, we propose multiple dataflow strategies towards varied CNN operators for efficient inference of MP-DNNs.}
\rFourCN{These dataflow strategies include FFCS for CONV operator, CF for PWCV operator, and FF for DWCV operator.}

\begin{figure}[tbp]
  \centering
    \includegraphics[width=0.98\columnwidth]{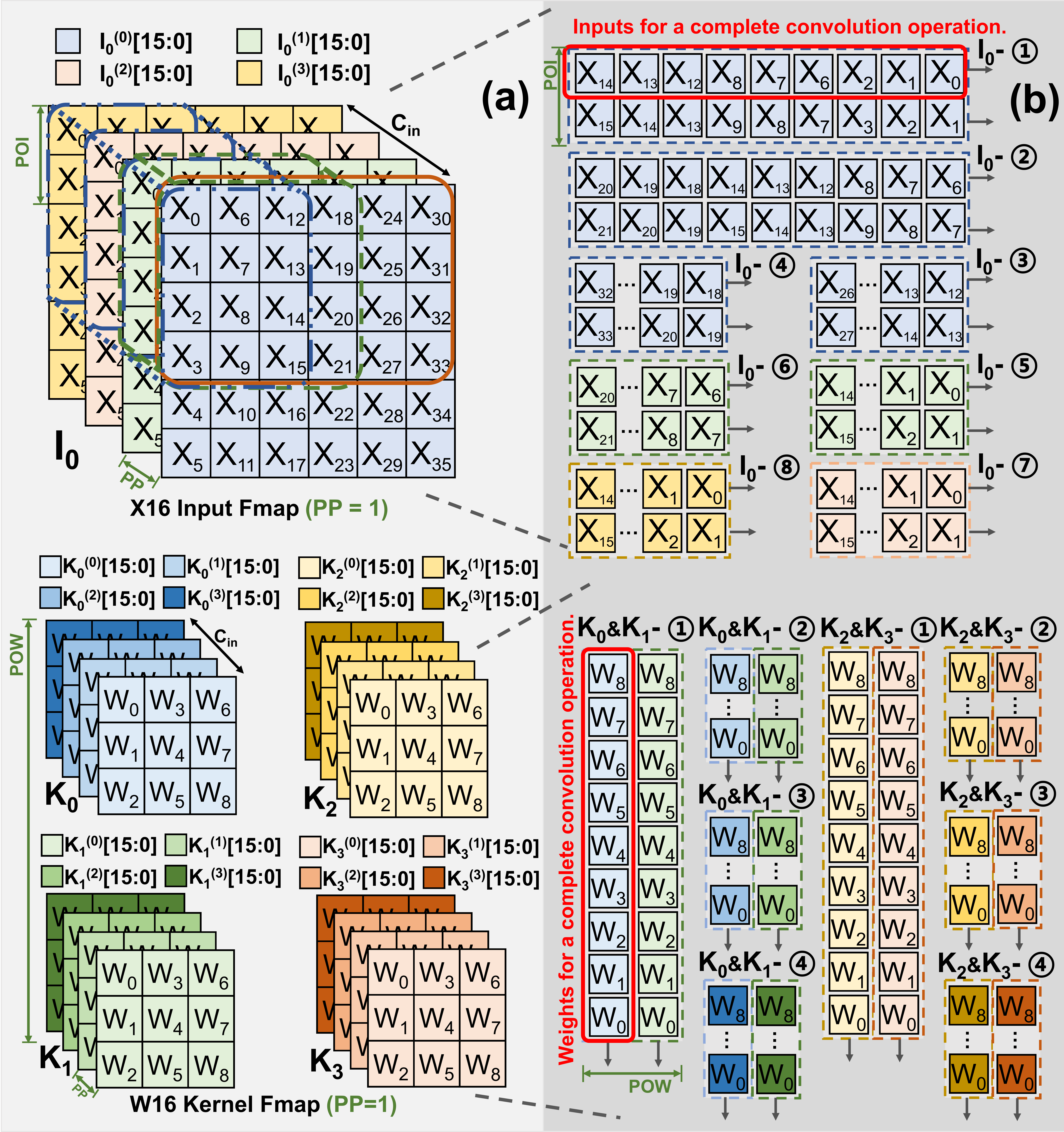}
      \caption{\reFirCN{Prefetching and partitioning of input Fmap and weight kernel Fmap: (a) data prefetch of input Fmap and weight kernel Fmap, (b) multiple sets of decomposed inputs and weights for complete convolution operations.}}
    \label{fig:conv_dataflow_1}
    \vspace{-0.8em}
\end{figure}

\WX{To provide a detailed explanation of the proposed convolution dataflow strategies, we take the four processing stages of 16-bit 3$\times$3 convolution computations as an example. 
\reFirCN{Fig.~\ref{fig:conv_dataflow_1} depicts the data pre-fetch of 16-bit inputs (X16) and 16-bit weights (W16).} 
In the case of 16-bit parallel computations, we set $POI$, $POW$, and $PP$ to 2, 2, and 1, respectively. 
In Fig.~\ref{fig:conv_dataflow_1}~(a), the upper part displays the pre-fetched data of the X16 input feature map (Fmap) for the four processing stages, with different dataflow strategies represented by the data inside the green dotted line (FFCS), blue dashed line (CF), and red solid line (FF). 
The lower part of Fig.~\ref{fig:conv_dataflow_1}~(a) illustrates the corresponding W16 kernel Fmap for four stages.}

\begin{figure*}[htbp]
  \centering
    \includegraphics[width=1.92\columnwidth]{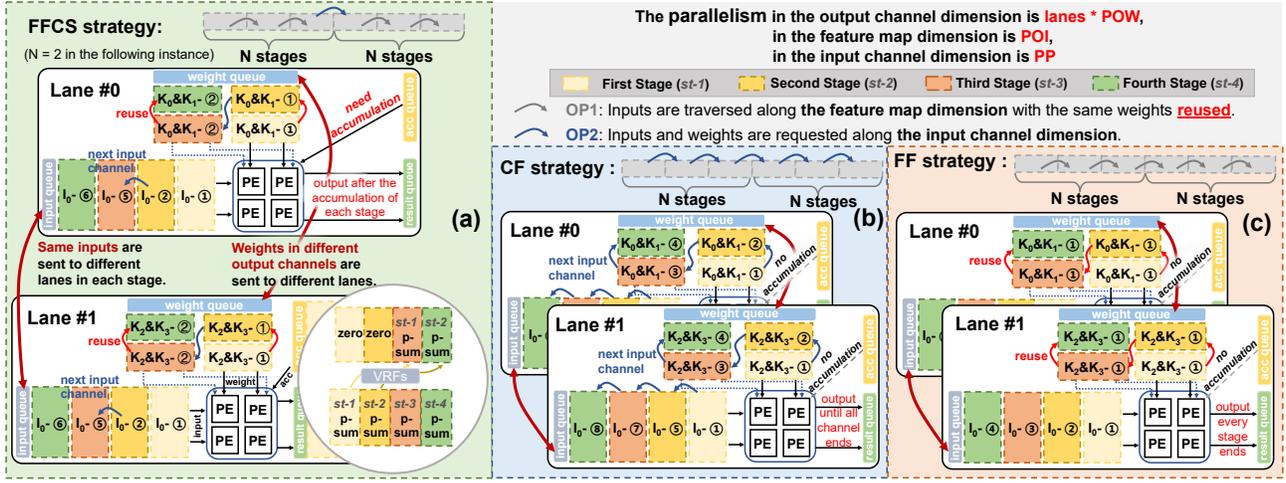}
      \caption{\reFirCN{Three convolution dataflow strategies of SPEED for supporting different convolutional operators: (a)-(c) are examples of computing dataflow of three convolution dataflow strategies, where the same inputs and the weights are sent to different lanes.}}
    \label{fig:conv_dataflow_2}
    \vspace{-0.8em}
\end{figure*}


\WX{Fig.~\ref{fig:conv_dataflow_1}~(b) illustrates multiple sets of inputs and weights for the FFCS, CF, and FF strategies during the convolution computation stages. 
Each set provides the operands required for parallel convolution computations. 
For instance, the inputs \ding{192}--\ding{195} are organized by traversing the 3$\times$3 convolution window in the first input channel of $I_{0}$, while inputs \ding{196}, \ding{197}, and \ding{198}, \ding{199} are organized from other input channels of $I_{0}$, respectively. 
Regarding the weights, $K_0$ and $K_1$ are combined along the input channel dimension to generate four sets of weights, and $K_2$ and $K_3$ follow a similar manner. 
Besides, as previously mentioned, three levels of parallelism computations are implemented in SPEED. 
The parallelism in the input channel dimension for inputs and weights is determined by $PP$, with $PP$ set as 1, 4, and 16 for 16-bit, 8-bit, and 4-bit operands, respectively. 
The parallelisms in the feature map and output channel dimensions are determined by $POI$ and $POW$, respectively.}

\WX{Fig.~\ref{fig:conv_dataflow_2} shows the data arrangement across stages for the three dataflow mapping strategies, where the operations can be classified into two categories: operation 1 (OP1) and operation 2 (OP2). 
In OP1, the inputs are traversed along the feature map dimension, utilizing the same weights as in the previous stage. 
In OP2, new inputs and weights are requested along the input channel dimension. 
The computation process for these strategies is elaborated as followings.}
\reFirCN{Additionally, in each stage, the same inputs are sent to two lanes, while weights of different output channels are distributed to specific lanes.}


\textbf{FFCS strategy.}
\WX{The data arrangement for the FFCS strategy involves multiple computation stages, denoted as ``N stages'' in Fig.~\ref{fig:conv_dataflow_2}~(a). 
Within these stages, SPEED executes OP1 to fully utilize the data reuse opportunities by multiplying inputs with the same weights.
After completing the N-th stage, SPEED performs OP2 to request data along the input channel dimension, relieving the storage pressure on VRFs.
In the subsequent N stages, SPEED repeats OP1 across stages, where the results of the latter N stages are accumulated with the partial sum obtained in the former N stages to reduce off-chip DRAM access.}

\WX{For example, in the case of N=2, inputs $I_{0}$--\ding{192} are multiplied with weights $K_{0}\&K_{1}$--\ding{192} and $K_{2}\&K_{3}$--\ding{192} in the first stage, while the weights are reused by $I_{0}$--\ding{193} in the next stage. 
In the third stage, inputs are traversed from the second input channel of $I_{0}$ and computed with the weights $K_{0}\&K_{1}$--\ding{193}, and $K_{2}\&K_{3}$--\ding{193}, respectively. 
Besides, the partial sums obtained in the first stage are requested from the VRFs and buffered in the accumulation queue, awaiting addition to the result of the current stage. 
Note that the partial sums of the third stage are stored in the same address as those of the first stage in the VRFs to minimize on-chip storage consumption. 
In the fourth stage, the weights of the previous stage are reused to perform convolution computations with $I_{0}$--\ding{197}. 
Subsequently, the current computation results are accumulated with the partial sums obtained in the second stage, and the results are stored back in the VRFs after the accumulation.}

\textbf{CF strategy.}
\jOneFC{PWCV operators are poorly suited for the FFCS strategy. The small convolution kernel size necessitates frequent VRF accesses for partial results, leading to significant data transmission overhead that dominates the overall computation time.}
\WX{To address this issue, SPEED employs the CF strategy to traverse along the input channel dimension firstly among stages, allowing for the accumulation of partial sums within the PE to eliminate extra data transmissions between MPTU and VRFs.
As depicted in Fig.~\ref{fig:conv_dataflow_2}~(b), in the first stage, the inputs from the first input channel, $I_{0}$--\ding{192}, are multiplied by weights $K_{0}\&K_{1}$--\ding{192} and $K_{2}\&K_{3}$--\ding{192}, where the results are stored in PEs. 
In the subsequent three stages, the inputs from the 2nd, 3rd, and 4th input channels of $I_{0}$ are performed with their corresponding weights. 
Besides, the computed results are consumed by adding to the partial sums previously stored in PE for the generation of a complete output, which continues until all input channels have been traversed. 
The final outputs are stored in the VRFs via the result queue.}


\textbf{FF strategy.}
\WX{For DWCV, since it decouples the dependency across different input channels, SPEED employs the FF strategy to improve computational efficiency without performing accumulation operations in the input channel dimension.
Specifically, the proposed FF strategy performs OP1 among different stages to fully exploit data reuse opportunities by traversing inputs within a single input channel with the same weights multiplied, which also reduces off-chip data accesses. 
As shown in Fig.~\ref{fig:conv_dataflow_2}~(c), inputs $I_{0}$--\ding{192} to $I_{0}$--\ding{195} are obtained by traversing along the first input channel of $I_{0}$, where each input performs computation with the same weights in every computation stage. 
The partial sum is sent to the VRFs through the result queue. 
The above operations are repeated in subsequent stages until the computations for a single input channel are completed.}

\begin{figure}[!tbp]
  \centering
    \includegraphics[width=1\columnwidth]{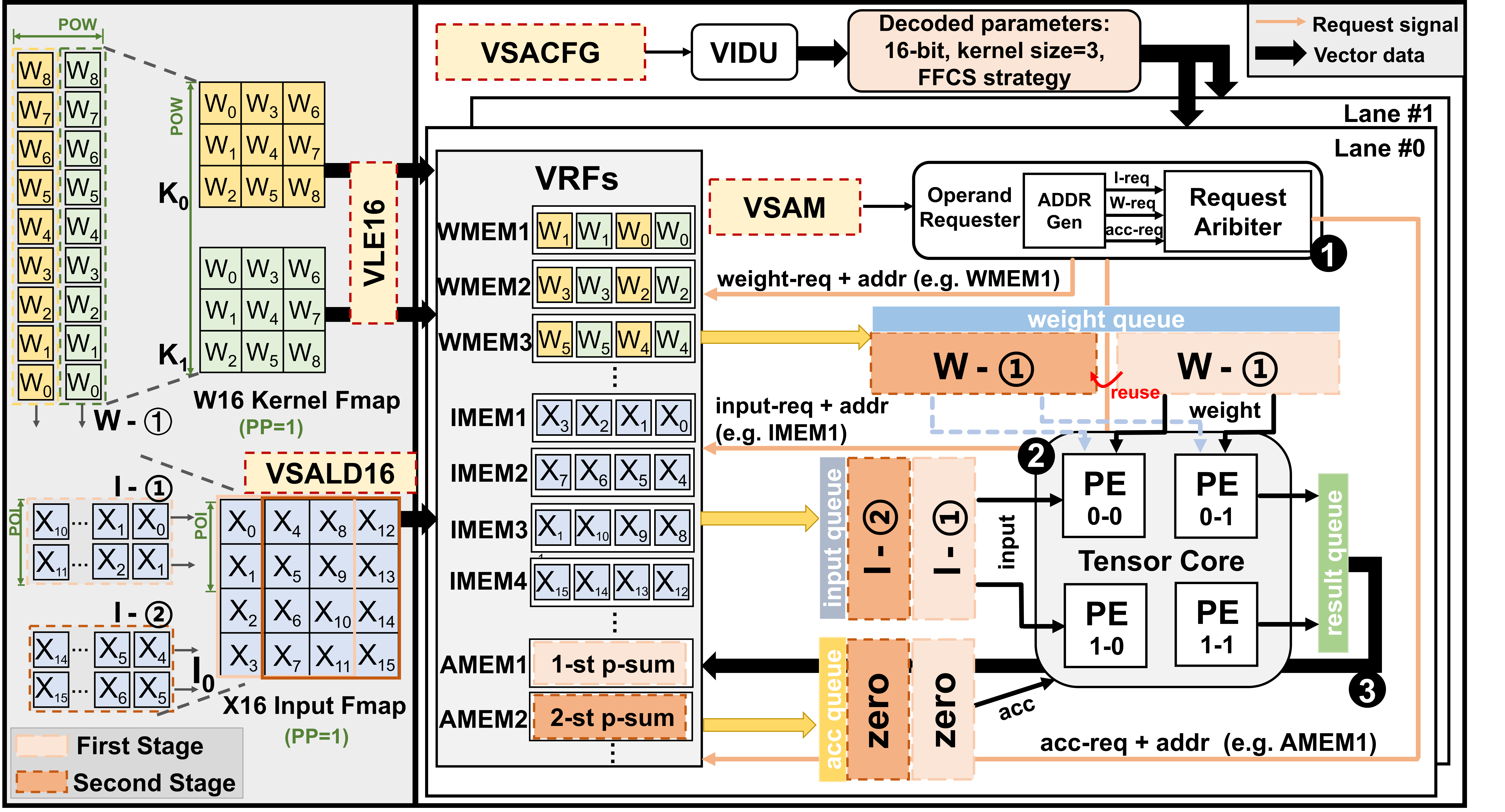}
      \caption{\reFirCN{An example of scheduling process in the first two stages of the FFCS dataflow strategy.}
       }
        \label{fig:dataflow_implement}
        \vspace{-10pt}
\end{figure}
\subsection{Execution Flow of Convolution Dataflow Strategy}\label{subsec: dataflow_implement}

\WX{To illustrate the scheduling process of dataflow strategies, we provide an example in Fig.~\ref{fig:dataflow_implement}, focusing on the first two stages of the FFCS strategy, where the hardware configurations remain consistent with those described in Sec.~\ref{subsec: conv_dataflow}. 
The scheduling process is controlled by a set of instructions comprising configuration-setting, memory access, and arithmetic instructions, as detailed in Sec.~\ref{subsec: instr_design}.}

\WX{Specifically, VIDU decodes the parameters within the \texttt{VSACFG} instruction, which are subsequently sent to the functional units for 16-bit 3$\times$3 convolution computations following the FFCS strategy. 
The memory access instructions, \texttt{VSALD} and \texttt{VLE}, are responsible for loading inputs and weights, such as $I_{0}$ and $K_0$\&$K_1$, into the VRFs, which buffer computing data for continuous computations.}

\WX{After completion of data loading, the \texttt{VSAM} instruction initiates the execution for convolution operations, where each customized arithmetic instruction enables performing operations across multiple stages. 
Specifically, each stage comprises three fundamental steps: \ding{202}~data-requesting; \ding{203}~computing; and \ding{204}~write-back to the VRFs. 
In the \ding{202}~data-requesting step, inputs, weights, and accumulation data are requested in different priorities from VRFs and then placed into their respective queues, which serve as temporary storage for the requested data. 
During the \ding{203}~computing step, inputs I--\ding{192} are multiplied by weights W--\ding{192}. 
Thanks to the subsequent addition between accumulation data and computed results, the request for accumulation data and multiplication computations can be executed concurrently, effectively overlapping the data-requesting with computing steps. 
In the \ding{204}~write-back step, the data of the result queue are written back to the VRFs. 
The concurrent accessibility of the three partitions within the VRFs allows the data-requesting step of the next stage to overlap with the current stage.}

\WX{Thus, in collaboration with the instructions and dataflow strategy, SPEED achieves efficient parameter configuration, data scheduling, and computation in MP-DNNs inference deployment. 
In particular, the customized instructions enable a high overlap when executing multi-stage operations, and the configurable multi-precision dataflow strategy fully leverages data reuse opportunities, efficiently enhancing computational and energy efficiency.}


%% file: 4-res-r1.tex
\pdfoutput=1

\section{Experimental Results}\label{sec:res}

\subsection{Experimental Setup}\label{subsec:setup}

\reFirCN{To demonstrate the effectiveness of the proposed SPEED processor, we first conduct a comprehensive evaluation of its computing performance.
Our benchmarking covers several MP-DNNs, with Ara~\cite{Cavalcante2022NewAra}, the pioneering open-source implementation of RVV processors, as the baseline for comparison.}
\reFirCN{
To ensure a fair comparison, SPEED and Ara are both configured with 4 lanes and a VRF size of 16KiB. 
Additionally, to match the computational resources of Ara, the MPTU of each lane is configured with \#TILE\_R and \#TILE\_C parameters, both equal to 2.
}

\jOneFC{To present SPEED's computational efficiency on MP-DNN workloads, we conduct benchmarks across a wide range of convolutional kernels and leading MP-DNN models that encompass a variety of MP-DNN operators, such as CONV, PWCV, DWCV, and MM. 
These models include VGG16\cite{Liu2015VGG16}, ResNet18\cite{He2016ResNet}, GoogLeNet\cite{Szegedy2015Googlenet}, MobileNetV2\cite{Sandler2018MobileNet}, ViT-Tiny\cite{Yuan2021ViT}, and ViT/B-16\cite{dosovitskiy2021an}, representing the diverse computational demands of contemporary MP-DNN applications.}
We employ \WX{\textit{performance \reFirCN{(ops/cycle)}}} as the evaluated metric, indicating the number of \WX{valid} computational results produced per clock cycle,
\reFirCN{where the clock count is measured by the cycle-accurate simulation with QuestaSim, following the experimental method in \cite{Cavalcante2022NewAra}.}
\reFirCN{
Furthermore, we synthesize SPEED using Synopsys Design Compiler 2022.03 on the TSMC 28 nm technology to explore the design space and analyze the trade-offs between throughput and area efficiency.}
The analysis provides insights into the scalability and adaptability of SPEED's architecture to different MP-DNN workloads.
Finally, we conduct a comprehensive comparison between our design and several state-of-the-art RISC-V processors\cite{Perotti2023Yun, Rossi2021VegaAT, XpulpNN2021TETC, Garofalo2023DARKSIDEAH, Dustin2023TCASI} \WX{to demonstrate the superior performance of SPEED in MP-DNN deployments.}

\subsection{Operator-level Evaluation}\label{sec:opera_level}


\jOneFC{This subsection delves into the operator-level evaluation, employing various dataflow strategies on SPEED and comparing them against Ara to demonstrate the effectiveness of the proposed mixed dataflow.}
\jOneFC{To ensure a fair comparison, SPEED and Ara are configured to achieve equivalent peak throughput \reFirCN{at 16-bit precision}, as elaborated in Sec.~\ref{subsec:setup}.}
\jOneFC{We first explore the efficient data reuse during computation under different dataflow strategies, as reflected in the \reFirCN{external memory access size,}} \WX{which is a key metric for evaluating energy and computational efficiency~\cite{Kang2022MOMM}.}
\jOneFC{In addition, performance serves as a direct measure of effective throughput on various DNN workloads.}
\jOneFC{Therefore, at the operator-level evaluation, we focus on exploring these key metrics to present the efficiency of SPEED under various DNN operators.}
\jOneFC{The benchmark DNN operators are PWCV, CONV$3$$\times$$3$, DWCV$3$$\times$$3$ (s = 2), and CONV$5$$\times$$5$, which are among the most common and computation-intensive operators in DNNs.}
Here, CONV$k$$\times$$k$ refers to the standard convolution kernel with a size of $k$, and DWCV$k$$\times$$k$ indicates the depth-wise convolution kernel, also of size $k$.
\reFirCN{Additionally, `s'  represents the stride length in convolution operations.}



\begin{figure}[!btp]
  \centering
    \includegraphics[width=0.92\columnwidth]{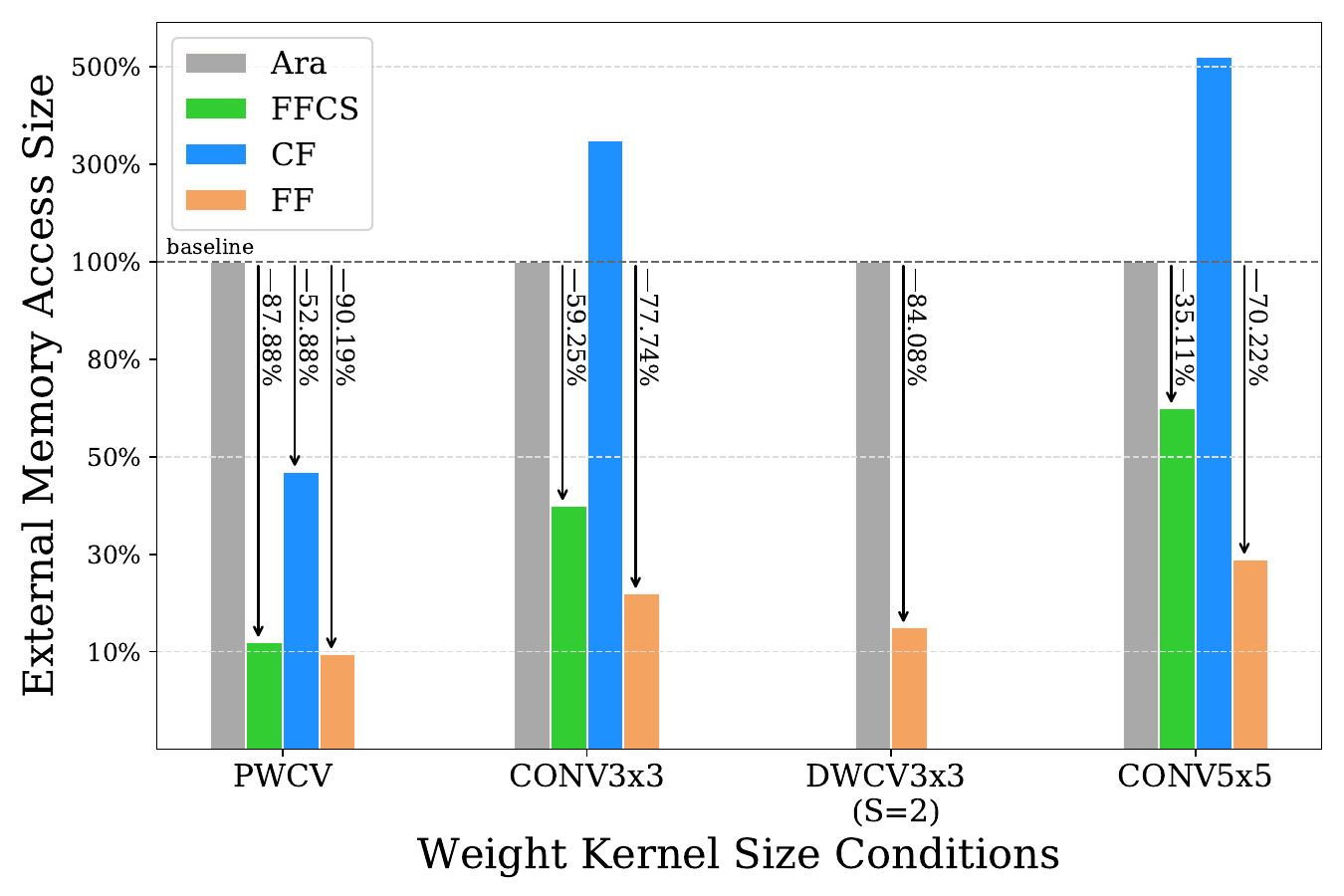}
      \caption{\reFirCN{External memory access size savings in SPEED that employs different dataflow strategies compared to Ara.}}
        \label{fig:access_size}
    \vspace{-10pt}
\end{figure}

\begin{figure*}[!tbp]
  \centering
    \includegraphics[width=1.9\columnwidth]{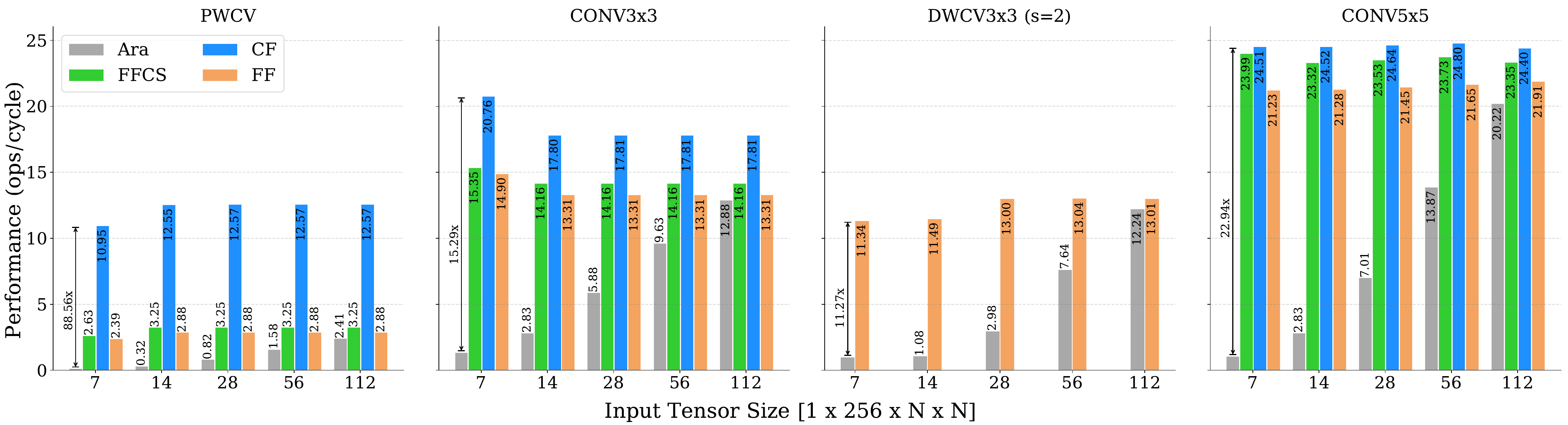}
    \vspace{-10pt}
    \caption{\reFirCN{Performance comparison of SPEED that employs different dataflow strategies with Ara across various DNN operator configurations.}}
    \label{fig:perf_kernel}
    \vspace{-1em}
\end{figure*}

\WX{Fig.~\ref{fig:access_size} exhibits various dataflow strategies in SPEED to effectively reuse data, significantly reducing off-chip \reFirCN{memory} access.}
\jOneFC{Specifically, in the evaluation of the PWCV operator, the \reFirCN{external memory access size} of SPEED using FFCS, CF, and FF dataflow strategies are merely 12.12\%, 47.12\%, and 9.81\% of Ara, respectively,} 
\reFirCN{leading to notable reductions in energy consumption and improved computational efficiency.}
\WX{Considering that both FFCS and CF strategies, developed for computations along the input channel dimension, are not applicable for the DWCV$3$$\times$$3$ operator, SPEED employs the FF strategy optimized for depth-wise operations, which achieves a \reFirCN{external memory access size} significantly reduced to 15.92\% of Ara’s.}
\rFourCN{
\jOneFC{Furthermore, based on the evaluation of the benchmark operators,}
\WX{SPEED employs the FF strategy for different operators to reduce \reFirCN{external memory access size} by 70.22\% to 90.19\% against Ara.}
Excluding the DWCV$3$$\times$$3$ operator, SPEED utilizing the FFCS strategy achieves \WX{significant savings in} \reFirCN{external memory access size} ranging from 35.11\% to 87.88\% compared to Ara.}

\jOneFC{Fig.~\ref{fig:perf_kernel} shows the evaluation of the performance of SPEED employing dataflow strategies in contrast with Ara under 16-bit precision DNN operators.}
\jOneFC{SPEED demonstrates consistently impressive inference performance across varying input tensor sizes, outperforming Ara, which experiences a significant drop in performance with smaller tensors.}
\reFirCN{The performance reduction of Ara is mainly due to its complex internal pipelined structure.}
\jOneFC{In terms of the PWCV operator, SPEED, employing the CF strategy, significantly outperforms Ara with performance improvements ranging from 5.21$\times$ to 88.56$\times$ across different input tensor sizes.}
Similarly, for \WX{the} DWCV$3$$\times$$3$ operator, the performance enhancement of the SPEED with FF strategy ranges from 1.06$\times$ to 11.27$\times$ compared to Ara.
\jOneFC{Under CONV$3$$\times$$3$ and CONV$5$$\times$$5$ operators, SPEED within the CF strategy exhibits performance improvements ranging from 1.38$\times$ to 15.29$\times$ and 1.21$\times$ to 22.94$\times$ over Ara, respectively.}

\jOneFC{Considering the \reFirCN{external memory access size} and performance metrics of SPEED with the three individual strategies, we can determine the most suitable strategies for different operator types.}
\jOneFC{As shown in Fig.~\ref{fig:access_size} and Fig.~\ref{fig:perf_kernel}, there is a trade-off between \reFirCN{external memory access size} and performance for the FFCS, CF, and FF strategies across various kernel sizes.}
\jOneFC{The CF strategy prioritizes performance compared to other proposed strategies and excels in PWCV operations due to its optimized data scheduling.
However, its high \reFirCN{external memory access size} can increase energy consumption.}
\jOneFC{In contrast, the FFCS strategy strikes a balance between performance and memory access for CONV operators. It maintains a lower \reFirCN{external memory access size} than Ara while achieving performance close to the CF strategy for larger kernels.}
\jOneFC{Finally, the FF strategy focuses on optimizing data reuse within the input feature map, leading to the lowest \reFirCN{external memory access size} among all strategies, making it particularly suitable for DWCV operations where each input channel is processed independently.}
\jOneFC{Based on the above analysis, our mixed dataflow scheduling includes CF, FFCS, and FF strategies for PWCV, CONV, and DWCV operations, respectively, which leverages the strengths of each strategy and achieves an optimal balance between performance, energy consumption, and memory access.}
\begin{figure}[!tbp]
  \centering
    \includegraphics[width=1\columnwidth]{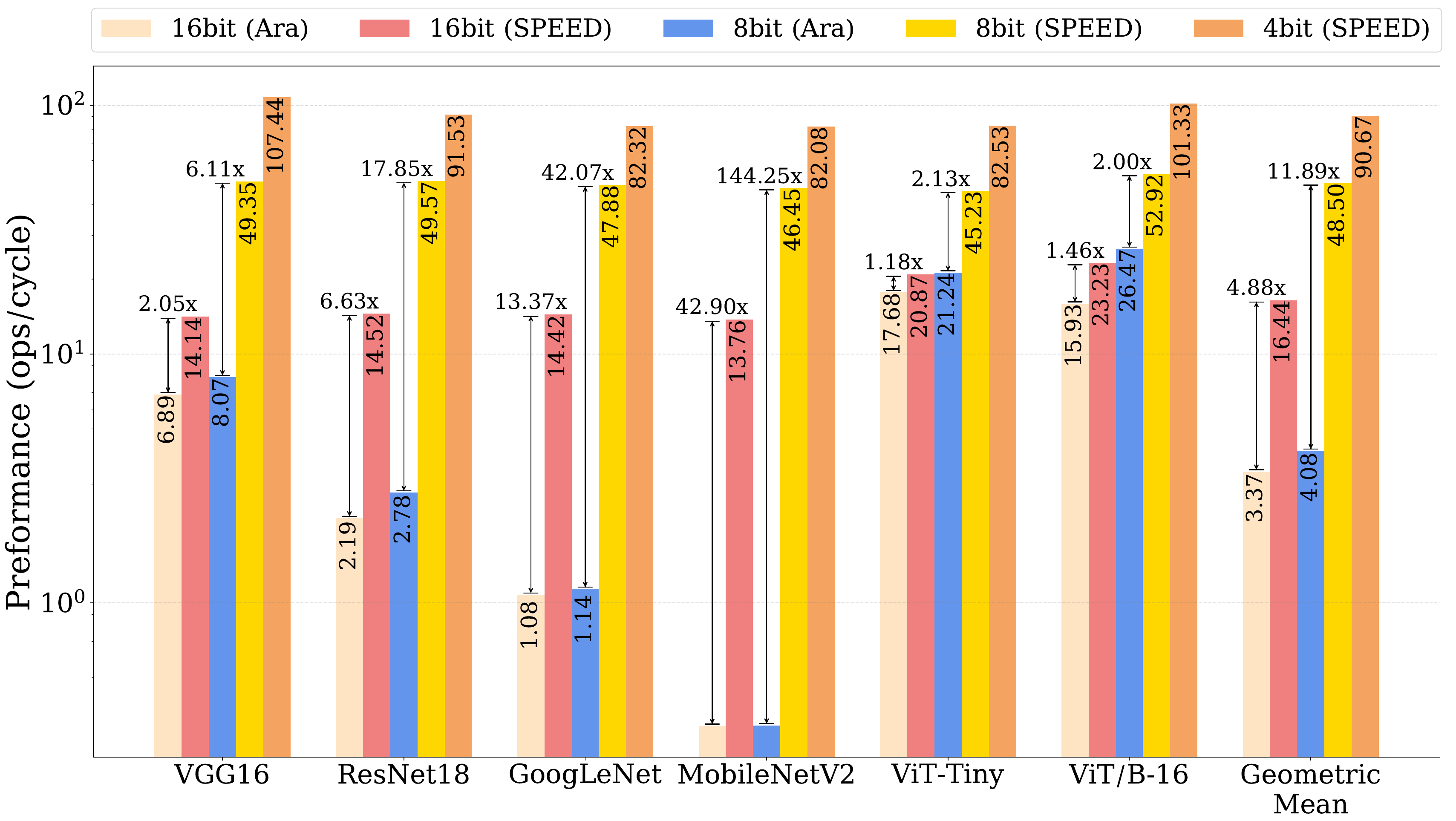}
    \vspace{-10pt}
    
      \caption{\reFirCN{Comparisons of the performance between Ara and SPEED. 
      On average, SPEED achieves a speedup of 4.88$\times$ and 11.89$\times$ over Ara at 16-bit and 8-bit precision, respectively.}
      }
    \vspace{-1em}
      
        \label{fig:model_level}
\end{figure}

\subsection{Model-level Evaluation}


\jOneFC{We conduct a model-level evaluation to present the performance benefits of SPEED that arise from the co-optimization of customized instructions, hardware architecture, and mixed dataflow strategies.}
\reFirCN{Four CNN-based and two Transformer-based DNN benchmarks, embracing VGG16, ResNet18, GoogLeNet, MobileNetV2, ViT-Tiny, and ViT/B-16, are evaluated within different convolution and MM layers under 16-bit, 8-bit and 4-bit precisions.}
\jOneFC{As shown in Fig.~\ref{fig:model_level}, SPEED achieves a 4.88$\times$ and 11.89$\times$ speedup over Ara on average at 16-bit and 8-bit precision, respectively.}

\rFourCN{For CNN-based models like VGG16, which heavily \WX{rely on} CONV$3$$\times$$3$ layers, SPEED achieves a 2.05$\times$ speedup over Ara at 16-bit precision, with advantages gained from the FFCS strategy.}
\jOneFC{Moreover, \WX{for} models dominated by PWCV and DWCV layers, such as ResNet18, GoogLeNet, and MobileNetV2, SPEED leverages its optimized CF and FF strategies to achieve significant speedups ranging from 6.63$\times$ to 42.90$\times$ at 16-bit precision and 17.85$\times$ to 144.25$\times$ at 8-bit precision, respectively.}
\jOneFC{For Transformer-based models like ViT-Tiny and ViT/B-16, SPEED \WX{employs a matrix multiplication dataflow strategy to deliver} a performance boost of 1.18$\times$ to 1.46$\times$ at 16-bit precision and 2.00$\times$ to 2.13$\times$ at 8-bit precision, respectively.}
In addition, SPEED enables efficient 4-bit inference with an average performance of up to 90.67 \reFirCN{ops/cycle}, surpassing the best of Ara by 22.22$\times$ on these DNN benchmarks.
\jOneFC{Notably, SPEED's performance at 8-bit and 4-bit precision is 2.95$\times$ and 5.51$\times$ that of 16-bit precision, respectively. 
}

\begin{table}[!tb]
\centering
\caption{ \reFirCN{Inference performance comparision between Ara and SPEED for various DNNs} } 
\renewcommand \arraystretch{1.6} 
\label{tab:comparison_full_net}
\resizebox{1\linewidth}{!}{
\begin{tabular}{ccccc}
\hline
Model
                             & \begin{tabular}[c]{@{}c@{}}Model size\\ FP32 / INT8 (MB)\end{tabular} & \begin{tabular}[c]{@{}c@{}}Top1 accuracy\\ FP32 / INT8\end{tabular} & \begin{tabular}[c]{@{}c@{}}Inference time\\ SPEED / Ara (cycle count)\end{tabular} & Relative  speedup \\ \hline
\multirow{2}{*}{VGG16}       & \multirow{2}{*}{553.4 / 138.4}                                       & \multirow{2}{*}{79.33 / 78.33}                                      & 622,010,560 / 3,677,525,600$^{\rm \dagger}$  & 6.11$\times$            
\\ 
                             &                                                                      &                                                                    
& 631,367,182 / 3,686,882,222$^{\rm \ast}$                                                      & 5.84$\times$            \\ \hline
\multirow{2}{*}{MobileNetv2} & \multirow{2}{*}{13.9 / 3.9}                                          & \multirow{2}{*}{78.00 / 77.33}                                      
& 13,395,597 / 1,932,019,408$^{\rm \dagger}$   & 144.25$\times$          \\
                             &                                                                      &                                                                     & 19,223,264 / 1,937,847,075$^{\rm \ast}$                                                         & 100.81$\times$          \\ \hline
\end{tabular}
}
\begin{tablenotes}
    \item[]\scriptsize $^{\rm \dagger}$ \reFirCN{Inference convolutional layers only.}
    \item[]\scriptsize $^{\rm \ast}$ \reFirCN{Complete application of specific DNN.}
\end{tablenotes}
\vspace{-10pt}
\end{table}


\reFirCN{
Futhermore, we select the VGG16 and MobileNetv2 DNN benchmarks, both quantized to 8-bit precision and evaluate complete application performance on the SPEED and Ara.
For effectively deploying these networks, the scalar processor manages floating-point operations and operations that are challenging to vectorize, such as max pooling. 
Additionally, vector processor performs convolution and easily vectorizable operations such as skip-connection.
As shown in the TABLE~\ref{tab:comparison_full_net}, in comparison to Ara, SPEED achieves a speedup of 6.11$\times$ in only convolution layers inference and 5.84$\times$ in the complete application condition for the VGG16 network. 
The predominant CONV layers benefit from SPEED's utilization of the FFCS strategy, resulting in a speedup several times greater than Ara. 
Similarly, for the MobileNetv2 network, SPEED demonstrates an improvement of 144.25$\times$ in only convolution layers inference and 100.81$\times$ in complete application performance, compared to Ara.
By leveraging the CF strategy to process the PWCV layers, which constitute a significant portion of MobileNetv2, SPEED achieves remarkable speedups up to several hundred times compared to Ara.
The speedup reduction between two inference methods due to the proportion of non-linear operations within the overall network inference duration is crucial in lightweight MobileNetv2 architecture.
These results highlight the performance benefits from the joint optimization of customized instructions, hardware architecture, and mixed dataflow strategies.
}

\begin{table}[!t]
\centering
\caption{ \reFirCN{Synthesize results of Ara and SPEED} } 
\renewcommand \arraystretch{1.18} 
\label{tab:comparison_per_lane}
\resizebox{0.8\linewidth}{!}{
\begin{tabular}{cccc}
\hline
\multirow{2}{*}{Parameter} & \multicolumn{2}{c}{Ara\cite{Cavalcante2022NewAra}} & \multirow{2}{*}{SPEED} \\
                           & reported~\cite{Askarihemmat2023QuarkAI}   & projected$^{\rm \ast}$  &                        \\ 
\hline
\multicolumn{1}{r}{Technology {[}nm{]}}    & 22          & 28         & 28                        \\
\multicolumn{1}{r}{Number of Lanes}        & 4           & 4          & 4                         \\
\multicolumn{1}{r}{VRF Size {[}KiB{]}}     & 16          & 16         & 16                        \\
\multicolumn{1}{r}{TT Frequency {[}GHz{]}} & 1.05        & 0.825      & 1.05                      \\
\multicolumn{1}{r}{Lane Area {[}mm2{]}}    & 1.20        & 1.94       & 1.08                      \\
\multicolumn{1}{r}{Lane Power {[}mW{]}}    & 229         & 229        & 71                  \\ \hline
\end{tabular}
}
\begin{tablenotes}
    \item[]\scriptsize $^{\rm \ast}$ \reFirCN{Projected from 22nm to 28nm assuming linear frequency scaling, quadratic area scaling, and constant power scaling (since Vdd does not scale).}
\end{tablenotes}
\vspace{-10pt}
\end{table}

\subsection{\reFirCN{Synthesis Result}}


\reFirCN{
On top of the operator-level and model-level evaluations, we implement the evaluated instance using TSMC 28nm technology, targeting an operating frequency of 1.05 GHz under typical corner (TT, 0.9V, \SI{25}{\degreeCelsius}).
To ensure a fair comparison, we select Ara with the same hardware configuration, as detailed in the paper~\cite{Askarihemmat2023QuarkAI}, including four lanes, 16 KiB of VRFs, and an equally configured scalar core, summarized in Table~\ref{tab:comparison_per_lane}.
Since the core computing unit, MPTU, is located in the vector processor, we only compare the synthesis results of a single lane for Ara and SPEED, where the results are projected to 28nm according to the method outlined in work~\cite{EIE2016ISCAHan}. 
As shown in TABLE~\ref{tab:comparison_per_lane}, the SPEED lane achieves a 45\% reduction in area and a 69\% decrease in power consumption than a standard Ara lane, primarily due to the float-point processing unit (FPU) removal and the incorporation of MPTU with high area efficiency and less power consumption. 
Considering the comparative results from operator-level and model-level evaluations, SPEED outperforms Ara across several MP-DNN operators and various MP-DNN benchmarks.
}

\begin{figure}[!tp]
    \centering
    \includegraphics[width=0.92\columnwidth]{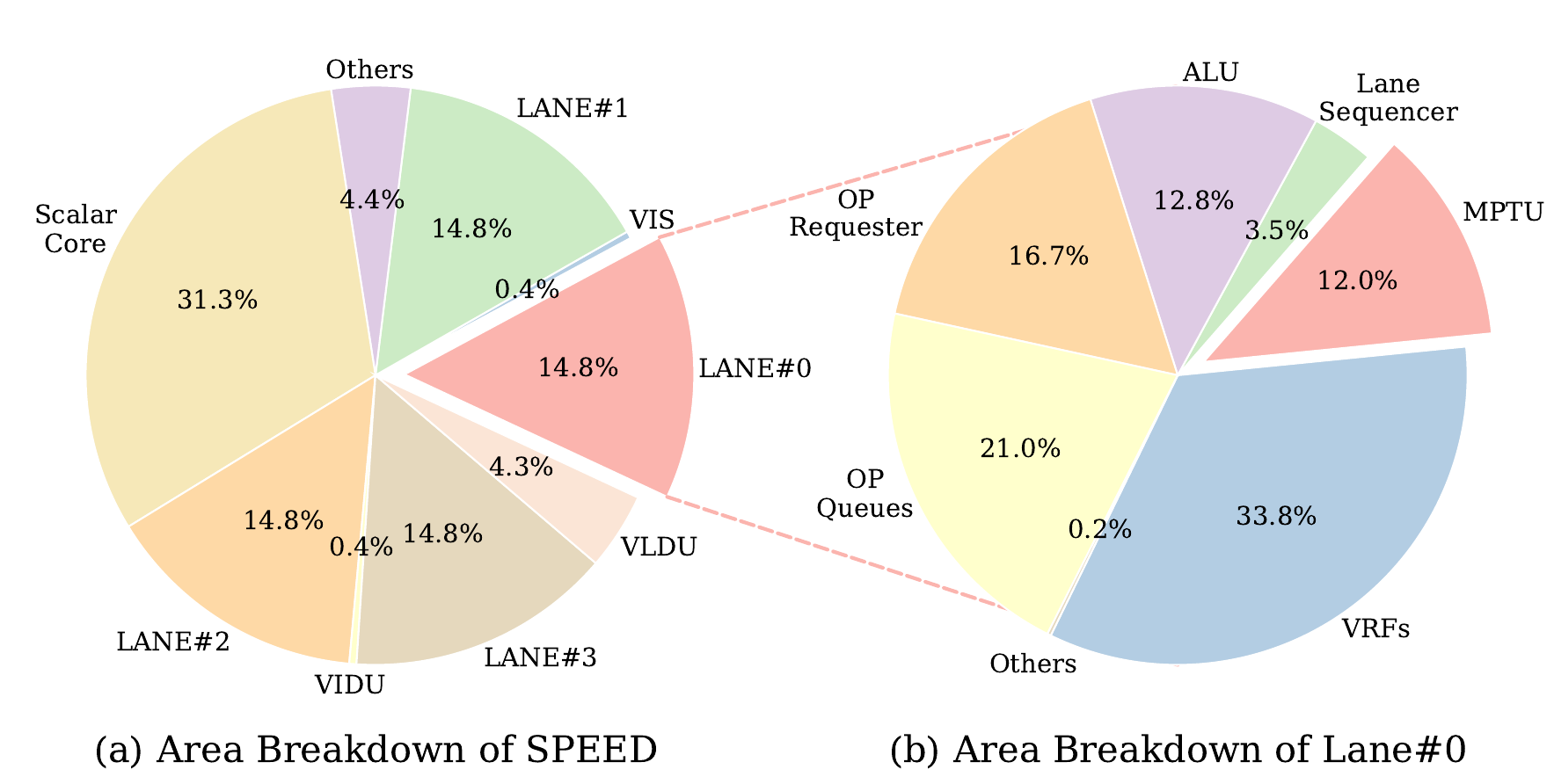}
        \caption{\reFirCN{Area breakdown of (a) SPEED and (b) a single lane. MPTU occupies only 12\% of the area in a single lane while achieving significant performance.}}
    \label{fig:area_breakdown}
\end{figure}

\reFirCN{Furthermore, we provide a detailed analysis of the area breakdown of SPEED with a scalar core, providing insights into the critical components contributing to the overall area.}
\jOneFC{Fig.~\ref{fig:area_breakdown}~(a) shows the area breakdown of SPEED.}
The majority of SPEED's area, up to \reFirCN{59\%}, is occupied by the lanes, which are responsible for executing the core computations with the specialized MPTU.
The remaining \reFirCN{41\%} is allocated to other components of the processor.

\jOneFC{A closer look at the lane's area allocation, as depicted in Fig.~\ref{fig:area_breakdown}~(b), reveals that the area of a lane is primarily consumed by four components: VRFs (\reFirCN{33\%}), OP Queues (\reFirCN{21\%}), OP Requester (\reFirCN{16\%}), ALU (13\%)  and MPTU (12\%).}
Over half the area is dedicated to data storage components like VRFs and OP Queues, \WX{whereas} computing units such as the ALU and MPTU only occupy 25\% of the area.
\jOneFC{Remarkably, the MPTU, occupying only 12\% of the lane area \WX{equivalent to \reFirCN{1.7\%} of the total area}, delivers a significant 22.22$\times$ performance improvement. 
This customized MPTU even consumes less area than the vector unit ALU, highlighting its area efficiency.}



\begin{figure}[!tbp]
  \centering
    \includegraphics[width=\columnwidth]{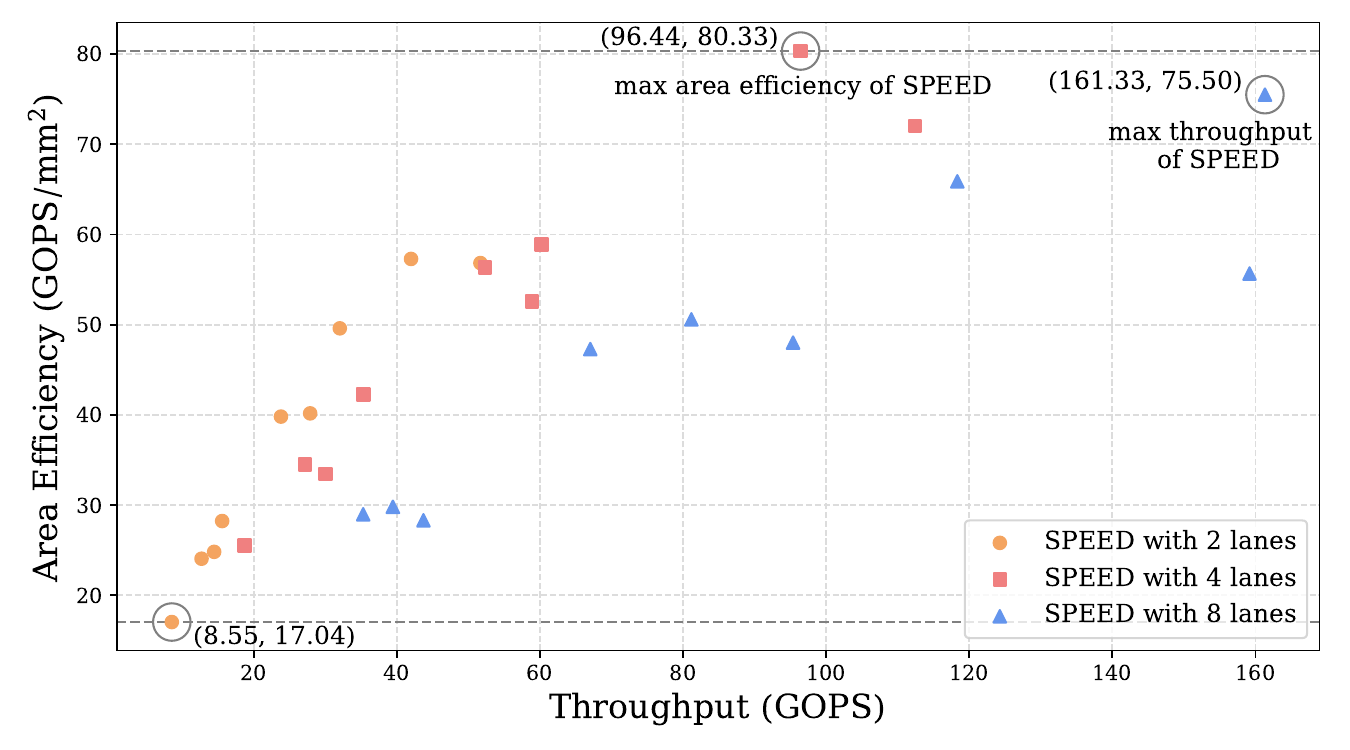}
      \caption{\reFirCN{Throughput and area efficiency comparison of SPEED with varying configurations.}}
        \label{fig:multi_lanes} 
    \vspace{-10pt}
\end{figure}

\begin{table*}[]
\centering
\caption{ \reFirCN{Comparison with the State-of-the-art RISC-V Processors} } 
\renewcommand \arraystretch{1.4} 
\label{tab:Related_Work_Summary}
\resizebox{0.96\textwidth}{!}{
\begin{tabular}{lccccccc}
\hline
& Yun~\cite{Perotti2023Yun} 
& Vega~\cite{Rossi2021VegaAT}                                                                  
& XPULPNN~\cite{XpulpNN2021TETC}                                                                
& DARKSIDE~\cite{Garofalo2023DARKSIDEAH}        
& Dustin~\cite{Dustin2023TCASI}                                                                                                                                  
& SPEED (Ours)$^{\rm \star}$                                                                    
\\ \hline
Technology   
& 65nm  
& 22nm        
& 22nm         
& 65nm                                                                    
& 65nm                    
& 28nm                                                                  
\\
Area                                                                       
& 6 $mm^2$             
& 12 $mm^2$                                                                
& 1.05 $mm^2$                                                               
& 12 $mm^2$          
& 10 $mm^2$                                                                 
& 1.20 $mm^2$                                                            
\\
INT Precision                                                              
& 8, 16, 32, 64 bit    
& 8, 16, 32 bit                                                              
& 2, 4, 8, 16, 32 bit                                                            
& 2, 4, 8, 16, 32 bit
& 2, 4, 8, 16, 32 bit                                                           
& 4, 8, 16, 32, 64 bit                                                              
\\ \hline
Supply Voltage                                                             
& 0.85 - 1.5 V              
& 0.5 - 0.8 V                                                              
& 0.6 - 0.8 V                                                            
& 0.75 - 1.2 V    
& 0.8 - 1.2 V                                                              
& 0.9 V                                                                       
\\
Max Frequency                                                              
& 280 MHz           
& 450 MHz                                                               
& 400 MHz                                                                
& 290 MHz         
& 205 MHz                                                                
& 1.05 GHz                                                                   
\\
Power Range                                                                
& \textit{\rm{N/A}}
& 1.7 uW - 49.4 mW                                                      
& 19.3 - 41.6 mW                                                         
& 213mW       
& 23 -156 mW                                                              
& 533 mW                                                     
\\ \hline
\begin{tabular}[c]{@{}l@{}}Best INT8 Performance (GOPS)$^{\rm \dagger}$\end{tabular}           
& 22.9 $\mid$ 53.2      
& 15.6 $\mid$ 12.3                                                            
& 23 $\mid$ 18.1                                                              
& 17 $\mid$ 39.4        
& 15 $\mid$ 34.8                                                                   
& 343.1
\\
\begin{tabular}[c]{@{}l@{}}Best INT8 Area Efficiency (GOPS/$mm^2$)$^{\rm \dagger}$ \end{tabular}       
& 3.8 $\mid$ 48.3 
& 1.3 $\mid$ 0.6                                                       
& 21.9 $\mid$ 10.6                                                         
& 1.4 $\mid$ 17.9    
& 1.5 $\mid$ 19.3                                                               
& 285.8
\\
\begin{tabular}[c]{@{}l@{}}Best INT8 Energy Efficiency (GOPS/W)$^{\rm \dagger}$\end{tabular}     
& 100.5 $\mid$ 233.3      
& 614 $\mid$ 482.4
& 1111 $\mid$ 872.9
& 191 $\mid$ 443.4       
& 303 $\mid$ 703.3
& 643

\\ \hline
\begin{tabular}[c]{@{}l@{}}Best Integer Performance (GOPS)$^{\rm \dagger}$\end{tabular}        
& 22.9 $\mid$ 53.2 (8b)    
& 15.6 $\mid$ 12.3 (8b)                                                        
& 72 $\mid$ 56.5 (2b)                                                           
& 65 $\mid$ 150.8 (2b)    
& 58 $\mid$ 134.6 (2b)                                                            
& 737.9 (4b)
\\
\begin{tabular}[c]{@{}l@{}}Best Integer Area Efficiency (GOPS/$mm^2$)$^{\rm \dagger}$\end{tabular}   
& 3.8 $\mid$ 48.3 (8b)     
& 1.3 $\mid$ 0.6 (8b)                                                         
& 68.5 $\mid$ 33.2 (2b)                                                         
& 5.4 $\mid$ 68.5 (2b)   
& 5.8 $\mid$ 74.7 (2b)                                                           
& 614.6 (4b)
\\
\begin{tabular}[c]{@{}l@{}}Best Integer Energy Efficiency (GOPS/W)$^{\rm \dagger}$\end{tabular} 
& 100.5 $\mid$ 233.3 (8b) 
& 614 $\mid$ 482.4 (8b)
& 3050 $\mid$ 2396.4 (2b)
& 835 $\mid$ 1938.4 (2b) 
& 1152 $\mid$ 2674.3 (2b)
& 1383.4 (4b)
\\ \hline
\end{tabular}
}
\begin{tablenotes}
    \item[]\scriptsize $^{\rm \star}$ \reFirCN{We configure SPEED with four lanes, and \#TILE\_R and \#TILE\_C parameters are set to 8 and 4, respectively, the highest area efficiency hardware condition.}
    \item[]\scriptsize $^{\rm \dagger}$ \reFirCN{Project from the reported technology to 28nm: linear, quadratic, and constant scaling for frequency, area, and power, respectively. Metrics are presented as ``reported $\mid$ projected''.}
\end{tablenotes}
\end{table*}

\subsection{Design Space Exploration of Scalable SPEED}\label{sec:design_space}

\jOneFC{Exploring the design space of SPEED is crucial for understanding its performance across diverse application scenarios and guiding optimal design choices for specific use cases.}
\jOneFC{We focus on two key metrics: area efficiency and throughput, which directly impact manufacturing cost and inference performance.}
\jOneFC{By delving into SPEED's design space, we can better grasp its strengths and limitations, providing valuable insights for system design and optimization.}

\reFirCN{SPEED supports three scalable module configurations, i.e., 2 lanes, 4 lanes, and 8 lanes. 
SPEED further offers nine available MPTU size configurations, with \#TILE\_R and \#TILE\_C parameters both configurable to 2, 4, or 8, which enlarges the scalable design space.}
\WX{Benefiting from its architectural advantages, SPEED provides a wide range of throughput options tailored to different application scenarios, 
\reFirCN{achieving a throughput from 8.5 GOPS to 161.3 GOPS under CONV3$\times$3 operator at 16-bit precision,}
as exhibited in Fig.~\ref{fig:multi_lanes}.}
\jOneFC{Meanwhile, SPEED 
\reFirCN{achieves a peak of 80.3 GOPS/$mm^2$ at a throughput of 96.4 GOPS,}  
demonstrating the benefits of joint optimization among customized instructions, hardware architecture, and dataflow mapping.}
\WX{Moreover, as the number of scalable modules in SPEED increases, it has a linear improvement in throughput with additional area overhead, which impacts the overall area efficiency.
As shown in Fig.~\ref{fig:multi_lanes}, \reFirCN{the instance of SPEED with 4 lanes achieves peak area efficiency, which efficiently balances processing throughput and area utilization.}}
\subsection{Comparisons with the State-of-the-Arts}\label{sec:related_works}


\reFirCN{We compare SPEED with the state-of-the-art RISC-V processors designed for DNN workloads, including Yun\cite{Perotti2023Yun}, Vega~\cite{Rossi2021VegaAT}, XPULPNN\cite{XpulpNN2021TETC}, DARKSIDE~\cite{Garofalo2023DARKSIDEAH} and Dustin~\cite{Dustin2023TCASI}. 
Table~\ref{tab:Related_Work_Summary} provides a comprehensive overview of these comparisons, focusing on projected area efficiency and energy efficiency under the same 28nm process node.}
\reFirCN{
Yun~\cite{Perotti2023Yun}, leveraging the official RVV extension ISA, has demonstrated high efficiency in DNN inference, attaining about 90\% FPU utilization in matrix multiplications\cite{Perotti2023Yun}.
Vega~\cite{Rossi2021VegaAT} is an fully programmable multi-core Internet-of-Things end-node SoC with legacy DNN acceleration capabilities in an advanced 22nm technology.
However, both lack native support for low-precision data operations crucial for quantized MP-DNNs.
}
\reFirCN{
XPULPNN~\cite{XpulpNN2021TETC}, DARKSIDE~\cite{Garofalo2023DARKSIDEAH}, and Dustin~\cite{Dustin2023TCASI} extend the already available RISC-V instructions to support low-precision computation, enhancing performance in sub-byte (e.g. 2-bit or 4-bit) through customized SIMD instructions and dedicated architecture.
SPEED extends this capability by supporting a broader range of precision formats and integrating high-throughput MPTU in vector lanes, enabling efficient inference for emerging MP-DNN models.}

\reFirCN{
{Compared} to Yun~\cite{Perotti2023Yun} and Vega~\cite{Rossi2021VegaAT}, SPEED shows significant throughput improvements, with 6.4$\times$ and 27.8$\times$ enhancements under 8-bit precision conditions, respectively. 
Regarding area efficiency, SPEED realizes a 5.9$\sim$12.7$\times$ improvement than Yun. 
Energy efficiency also sees notable improvements, with SPEED achieving 2.7$\sim$5.9$\times$ and 1.3$\sim$2.8$\times$ enhancements in contrast with Yun and Vega, respectively.
Contrary to Yun and Vega, thanks to the customized RVV instructions, SPEED supports lower-precision (than 8-bit) integer workloads and enables the computation of emerging DNN models that employ quantization schemes~\cite{Wang2019CVPR, zhou2018explicit}, further improving throughput and area efficiency under best integer conditions.
{Compared} to XPULPNN~\cite{XpulpNN2021TETC}, SPEED exhibits a gain of up to 18.9$\times$ and 26.9$\times$ under 8-bit precision regarding throughput and area efficiency, respectively.
Moreover, SPEED enables efficient inference that surpasses the best of XPULPNN by 13.1$\times$ and 18.5$\times$, respectively, on integer performance and area efficiency. 
{Compared} to the DARKSIDE~\cite{Garofalo2023DARKSIDEAH} and Dustin~\cite{Dustin2023TCASI}, SPEED shows significant throughput improvements, with speedups of 8.7$\times$ and 9.8$\times$ under 8-bit precision and 4.8$\times$ and 5.4$\times$ under best integer condition, respectively.
Additionally, SPEED also demonstrates an enhancement of 15.9$\times$ and 14.8$\times$ for 8-bit precision, 8.9$\times$, and 8.2$\times$ for best integer condition in terms of area efficiency.
Regarding energy efficiency, XPULPNN stands out due to low voltage (i.e., down to 0.6V) operating mode and cluster architecture with a simple core-memory traffic scheme. 
DARKSIDE and Dustin deliver reduced power consumption compared to SPEED because of their focused low-power execution mode and specific engines.
However, these works enable multi-precision computations by combining single-precision computing units and perform matrix and convolution operators with respective engines, further resulting in insufficient area efficiency.
Contrarily to works~\cite{XpulpNN2021TETC, Garofalo2023DARKSIDEAH, Dustin2023TCASI}, SPEED offers up to 26.9$\times$ better area efficiency, with a considerable throughput performance gain of up to 18.9$\times$, thanks to the multi-precision processing capability of MPTU and the proposed mixed dataflow strategy operating in MPTU, achieving significant performance improvements within limited area constraints.
In summary, SPEED significantly outperforms the mentioned RISC-V works~\cite{Perotti2023Yun, Rossi2021VegaAT, XpulpNN2021TETC, Garofalo2023DARKSIDEAH, Dustin2023TCASI}, achieving throughput improvements of up to 13.8$\times$ at best integer condition.
Additionally, SPEED demonstrates remarkable area efficiency, with enhancements ranging from 8.2$\times$ to 18.5$\times$ for MP-DNN inference under the constrained area conditions typical in edge computing scenarios.
}

%% file: 5-concls.tex
\pdfoutput=1
\section{Conclusion}\label{sec:concls}
\WX{In this paper, we propose SPEED, a scalable RISC-V vector (RVV) processor designed to facilitate efficient multi-precision DNN (MP-DNN) inference. 
Specifically, we introduce several customized RVV instructions tailored to support MP-DNN operations ranging from 4-bit to 16-bit precision, reducing the instruction complexity involved in the MP-DNN inference tasks. 
To enhance the parallel processing capabilities of the scalable modules, we design a multi-precision tensor unit in SPEED that offers configurable parallelism that matches various MP-DNN operators.}
\jOneFC{Moreover, we employ a flexible mixed dataflow method devised specifically for various MP-DNN operators, further improving computational efficiency.}
\reFirCN{
Experimental results demonstrate that SPEED achieves 4.8$\sim$13.8$\times$ higher throughput, as well as 8.2$\sim$18.5$\times$ higher area efficiency over the prior arts for best integer performance.
}